\documentclass[preprints,article,accept,pdftex,moreauthors]{Definitions/mdpi} 
\usepackage{mathtools}
\usepackage{tensor}
\usepackage{tikz,pgfplots} %tikz for figures in Latex
\usepackage{tikz-cd}
\pgfplotsset{compat=1.18}
\usetikzlibrary{decorations.pathreplacing} %for more fancy stuff in tikz
\usetikzlibrary{patterns}
\usetikzlibrary{calc}
\usetikzlibrary{arrows.meta,arrows,positioning}

\usepackage{csquotes}
\usepackage{amsmath,amsfonts}
\usepackage{amssymb,latexsym}
\usepackage{color}
\usepackage{graphicx}

\usepackage[normalem]{ulem} %for crossing out text

\newcommand{\ii}{\mathrm{i}}
\newcommand{\ud}{\mathrm{d}}

\newcommand{\uD}{\mathrm{D}}
\newcommand{\uM}{\mathrm{M}}
\newcommand{\uE}{\mathrm{E}}

\firstpage{1} 
\pubvolume{1}
\issuenum{1}
\articlenumber{0}
\pubyear{2025}
\copyrightyear{2025}
\datereceived{ } 
\daterevised{ } % Comment out if no revised date
\dateaccepted{ } 
\datepublished{ } 
\hreflink{https://doi.org/} % If needed use \linebreak
\Title{
Torsion-Induced Quantum Fluctuations in Metric-Affine Gravity using the Stochastic Variational Method
}

\TitleCitation{Title}

\Author{Tomoi Koide $^{1,\dagger,\ddagger}$\orcidA{} and Armin van de Venn $^{2,3,\ddagger}$\orcidB{}}

\AuthorNames{T. Koide and A. van de Venn}

\isAPAStyle{%
       \AuthorCitation{Koide T., \& van de Venn, A.}
         }{%
        \isChicagoStyle{%
        \AuthorCitation{Koide T., and van de Venn, A.}
        }{
        \AuthorCitation{Koide T., ; van de Venn, A.}
        }
}

\address{%
$^{1}$ \quad Instituto de F\'{\i}sica, Universidade Federal do Rio de Janeiro, 
21941-972, Rio de Janeiro, RJ, Brazil; tomoikoide@gmail.com,koide@if.ufrj.br\\
$^{2}$ \quad Frankfurt Institute for Advanced Studies (FIAS), Ruth-Moufang-Str.~1, 60438 Frankfurt am Main, Germany; venn@fias.uni-frankfurt.de\\
$^{3}$ \quad Physics Department, Goethe University, Max-von-Laue-Str.~1, 60438 Frankfurt am Main, Germany}

\secondnote{These authors contributed equally to this work.}
\abstract{
This review paper comprehensively examines the influence of spatial torsion on quantum fluctuations from the perspectives of Metric-Affine Gravity (MAG) and the Stochastic Variational Method (SVM). We first outline the fundamental framework of MAG, a generalized theory that includes both torsion and non-metricity, and discuss the geometrical significance of torsion within this context. Subsequently, we summarize SVM, a powerful technique that facilitates quantization while effectively incorporating geometrical effects. By integrating these frameworks, we evaluate how the geometrical structures originating from torsion affect quantum fluctuations, demonstrating that they induce non-linearity in quantum mechanics. Notably, torsion, traditionally believed to influence only spin degrees of freedom, can also affect spinless degrees of freedom via quantum fluctuations. Furthermore, extending beyond the results of previous work [Koide and van de Venn, Phys. Rev. A112, 052217 (2025)], we investigate the competitive interplay between the Levi-Civita curvature and torsion within the non-linearity of the Schrödinger equation. Finally, we discuss the structural parallelism between SVM and information geometry, highlighting that the splitting of time derivatives in stochastic processes corresponds to the dual connections in statistical manifolds. These insights pave the way for future extensions to gravity theories involving non-metricity and are expected to deepen our understanding of unresolved cosmological problems.
}


\begin{document}

%%%%%%%%%%%%%%%%%%%%%%%%%%%%%%%%%%%%%%%%%%

\section{Introduction}
Einstein's theory of general relativity (GR) has achieved remarkable success 
in describing gravitational phenomena on solar system scales. 
However, our understanding of gravity is still far from complete. 
It is well-known that there are four fundamental forces in nature: the strong, weak, electromagnetic, 
and gravitational interactions. The first three of these are governed by the laws of quantum mechanics. 
It is widely regarded as unlikely that gravity would be the sole exception, and despite numerous attempts to quantize it, a definitive theory has not yet been established.
Furthermore, puzzling observational 
results concerning gravitational interactions have been reported. The accelerated expansion of the universe, 
attributed to "dark energy," and the anomalous rotation of galaxies, suggesting the presence of "dark matter," 
can be seen as indications of the potential need for an extension of standard GR.

Metric-Affine Gravity (MAG) has emerged as a promising candidate for such a generalized theory \cite{Hehl95, Hehl99, Baldazzi21, Francois25, Blagojevic12, Obukhov06, Mielke17}. 
The idea of relaxing the constraints of Riemannian geometry is not new, 
tracing its historical roots to early attempts by Weyl, Cartan, and Einstein to unify gravity with 
other forces or to incorporate the intrinsic spin of matter into the geometric fabric of spacetime \cite{Weyl1918, Cartan1922, Einstein1925}. 
Unlike GR, which is based on Riemannian geometry where the connection is uniquely determined by the metric 
(as the Levi-Civita connection), 
MAG treats the metric and the affine connection as independent geometric entities. 
This generalization inevitably introduces two new geometric properties: torsion and non-metricity. 
Torsion, the antisymmetric part of the connection, relates to the non-commutativity of parallel transport, 
while non-metricity implies that the length of a vector is not conserved under it. 
The attempt to explain cosmological puzzles without postulating dark components 
by utilizing this richer geometric framework is an intriguing one \cite{vandeVenn2022, Kirsch2023, Benisty2021,Pereira:2022,Milton:2020, Bengochea2008, Rigouzzo2023, Megier2020, Poplawski2011, Gonner84, Minkevich07, Shie08, Gasperini86, Trautman73, Cai15, Obukhov87}.

Let us consider the contribution of torsion in such a generalization. For minimal coupling of scalar fields and gauge vector fields to curved spacetime, the affine connection either does not appear at all or enters only through combinations fixed by gauge invariance that are insensitive to torsion. Consequently, at the level of minimal coupling, the classical equations of motion for such fields are unaffected by spacetime torsion.
In contrast, fermionic fields necessarily require a local Lorentz frame for their definition. Spinor representations exist only with respect to the local Lorentz group, and the corresponding Dirac equation in curved spacetime involves gamma matrices defined in a local inertial frame. The vielbein provides the map between this local frame and spacetime coordinates, while the affine/spin connection enters the fermionic covariant derivative. Since torsion contributes directly to the antisymmetric part of the spin connection, fermions couple to torsion already at the level of minimal coupling.

Due to this historical context, the prevailing view has long been that the effects of spacetime torsion can only be observed through the behavior of fermionic particles. 
At the same time, modified gravity theories based on torsion have recently been extensively applied to cosmological contexts, providing alternative explanations for the accelerated expansion of the universe and being actively tested against observational data \cite{PhysRevD.97.084008, Carloni2025}. 
Motivated by this growing interest, the primary purpose of this review is to urge a fundamental reconsideration of the established view of torsion-matter coupling. We employ SVM as a powerful tool to demonstrate a novel quantum effect: contrary to the classical expectation, torsion can couple even to spinless particles through their quantum fluctuations. Consequently, at the quantum level, the effects of torsion may manifest universally.
To understand this rigorously, it is necessary to use a quantization method that naturally incorporates geometric effects, namely the Stochastic Variational Method (SVM) \cite{Koide2015,Koide2016}.

Historically, quantum mechanics is obtained by applying a procedure known as canonical quantization to 
the corresponding classical equations. 
This procedure requires that when replacing the canonical variables of classical mechanics 
with their corresponding self-adjoint operators, the pair of canonical variables must 
satisfy canonical commutation relations instead of Poisson brackets. While there are exceptions, 
this method works remarkably well in flat spacetime and for cases like Cartesian coordinates 
where the spectrum of the position operator is not bounded. However, if the spacetime is not flat, 
there is no guarantee that the framework of canonical quantization will succeed. 
Indeed, it is known that this procedure cannot be applied in cases like polar coordinates where 
the spectrum of the position operator is bounded. 
The well-known paradox concerning the uncertainty relation 
for angular variables is related to this limitation of canonical quantization. 
See, for example, Ref.\ \cite{U_Gazeau_2020} and references therein. 
The fact that difficulties arise even in such simple coordinate systems suggests 
the formidable challenge of applying quantization 
to classical systems interacting with GR, 
where spacetime itself is dynamic, or to MAG with its even more complex geometry.

To overcome this limitation of canonical quantization, various alternative approaches have been proposed 
for the quantization of systems with non-trivial topologies. Among them, 
the affine covariant integral quantization, developed by J.-P. Gazeau to whom this special issue is dedicated, 
provides a powerful framework \cite{romain2017,Romain2022,bergeron2024,romain2025}. 
Furthermore, approaches that regard quantum fluctuations as a stochastic process have been explored 
in various ways since the original idea of F\"{u}rth and Feyn\'{e}s 
\cite{furth1933,feynes1952}. 
For example, the stochastic quantization of Parisi and Wu reproduces quantum theory 
as the stationary state of a stochastic process over a fictitious time variable \cite{parisi1981}.

Among these important approaches, this paper focuses on SVM proposed 
by Yasue \cite{S_Yasue_1981}. This method, a reformulation of Nelson's stochastic mechanics \cite{D_Nelson_1966} as a variational principle, 
is characterized by its direct derivation of dynamics from the physical action by utilizing a variational method, 
viewing quantum fluctuations as a fundamental stochastic process. 
The SVM's formulation is inherently geometric, allowing it to naturally handle systems with 
non-trivial topologies and bounded spectra, where canonical quantization faces difficulties, 
in a coordinate-independent manner \cite{N_Koide_2019,U_Gazeau_2020,V_Koide_2020}. Another significant advantage is that it provides 
a physically intuitive picture, such as explaining the uncertainty principle through the non-differentiability 
of the particle's trajectory in Brownian motion \cite{U_Gazeau_2020,G_Koide_2018,U_Goncalves_2020,Koide_2022_JSTAT}.

Note that, in this review, we primarily focus on the effects of spatial torsion, which is treated strictly as a fixed background field, to elucidate its impact on quantum fluctuations. 
Consequently, our current approach relies on a non-relativistic formulation. 
While a fully metric-affine framework fundamentally demands Lorentz covariance, introducing a stochastic process in a relativistic setting encounters profound theoretical challenges. 
For instance, formulating a manifestly covariant Wiener process (noise) remains an open problem. 
Furthermore, as is well known from the historical Newton-Wigner problem \cite{RevModPhys.21.400}, defining a classical-like trajectory for a single particle in relativistic quantum mechanics is fundamentally difficult due to particle creation and annihilation processes. 
Since the present SVM framework incorporates quantum effects via the fluctuation of particle trajectories, avoiding these profound field-theoretical complications necessitates restricting our current scope to non-relativistic particle motion. 
A comprehensive covariant extension, which likely requires transitioning to the stochastic quantization of fields (see, e.g., Ref.\ \cite{Koide2015}), as well as the full inclusion of dynamical spacetime torsion, are left as major subjects for future work.

A secondary objective is to discuss the structural parallels between SVM and information geometry. This discussion is motivated by recent developments in $f(Q)$ (or Symmetric Teleparallel) Gravity \cite{nester1999,beltran2019,heisenberg2024}, which describes gravity solely through non-metricity. While Iosifidis and Pallikaris \cite{iosifidis2023} have touched upon the connection between such generalized gravity theories and information geometry, this intersection remains largely uncharted territory. In this review, we provide a brief introduction to information geometry and elucidate its structural similarity to SVM, paving the way for extending our methods to a more general framework that incorporates the effects of non-metricity in future research.

This paper is organized as follows. 
In Sec. \ref{sec:math_mag}, we review the mathematical foundations of gauge theories, with particular emphasis on their application to MAG, followed by the definitions of its key geometric quantities, curvature, torsion, and non-metricity. 
Sections \ref{sec:clas} through \ref{sec:numerical} serve as a comprehensive review of the established foundations of SVM.
Specifically, after setting the stage with the classical variational principle in Sec.\ \ref{sec:clas}, we introduce the stochastic formalism in Sec.\ \ref{sec:svm_general}, derive the Schr\"{o}dinger equation in Sec.\ \ref{sec:svm_particle}, and establish the stochastic Noether theorem in Sec.\ \ref{sec:s-noether}. 
Section \ref{sec:numerical} presents a numerical validation of this framework by simulating the double-slit experiment. 
In contrast to these review sections, Sections \ref{sec:svm_torsion} and \ref{sec:svm_torsion2} present our genuinely new contributions.
The extension of SVM to curved backgrounds with torsion is discussed in Sec.\ \ref{sec:svm_torsion}. 
Section \ref{sec:svm_torsion2} then presents the derivation of the novel non-linear Schr\"{o}dinger equation, where the non-linearity is governed by the competitive interplay between the Levi-Civita curvature and the torsion scalar, and discusses its physical implications and cosmological constraints. 
Section \ref{sec:inf_geo} introduces the basics of information geometry and discusses a new structural parallelism with SVM. 
Section \ref{sec:summary} is devoted to a summary and concluding remarks.

\section{Mathematical Foundations of Metric-Affine Gravity} \label{sec:math_mag}
Gauge theories are naturally formulated in the language of principal and associated bundles. While Yang–Mills theories are well understood within this framework, we now aim to review an analogous geometric description for the case of MAG.

\subsection{Frame Bundle and Gauge Structure}
Let $\mathcal{M}$ be a smooth 4-dimensional Lorentzian manifold.
Within the context of MAG, the relevant principal bundle is the frame bundle $F\mathcal{M}$ over $\mathcal{M}$ with structure group $\mathrm{GL}(4,\mathbb{R})$. As a set, $F\mathcal{M}$ is defined as
\begin{equation}
    F\mathcal{M} \coloneqq \bigsqcup_{p\in \mathcal{M}} \mathrm{Iso}(\mathbb{R}^4,T_p\mathcal{M}),
\end{equation}
where $\mathrm{Iso}(\mathbb{R}^4,T_p\mathcal{M})$ denotes the set of all 
vector space isomorphisms $f_p\colon \mathbb{R}^4 \to T_p\mathcal{M}$
and $\bigsqcup$ denotes the disjoint union. 
An isomorphism between $\mathbb{R}^4$ and $T_p\mathcal{M}$ is 
equivalent to the choice of an ordered basis or \textit{frame} in $T_p\mathcal{M}$. 
Hence the name frame bundle.
Any element of $F\mathcal{M}$ admits the form $(p,f_p)$, where $p\in \mathcal{M}$
and $f_p \in \mathrm{Iso}(\mathbb{R}^4,T_p\mathcal{M})$. The group 
$\mathrm{GL}(4,\mathbb{R})$ acts from the right on such an element 
as
\begin{equation}
    (p,f_p)\cdot g \coloneqq (p,f_p \circ \rho(g)) \in F\mathcal{M}
\end{equation}
for any $g \in \mathrm{GL}(4,\mathbb{R})$. Here,
$\rho \colon \mathrm{GL}(4,\mathbb{R}) \to \mathrm{Aut}(\mathbb{R}^4)$ is the standard representation associating to every invertible $4\times 4$ matrix its corresponding automorphism on $\mathbb{R}^4$.
Intuitively, composing $f_p$
with $\rho(g)$ provides a different choice of frame.

The notion of gauge fields in this framework is understood in terms of the so-called \enquote{connection 1-forms} on a principal bundle. They give rise to covariant derivatives on the associated bundles.
In general, a connection 1-form $\omega$ on a principal $G$-bundle $H$ is an element of $\Omega^1(H,\mathfrak{g})$ satisfying the usual equivariance and verticality conditions, where $\mathfrak{g}$ is the Lie Algebra associated to the Lie Group $G$. Here, $\Omega^1(H,\mathfrak{g})$ denotes the set of 1-forms on $H$ taking values in $\mathfrak{g}$. In other words, $\omega(p)$ is a linear map $T_p H \to \mathfrak{g}$ for any $p\in H$.
In our specific case the connection $\omega$ is a $\mathfrak{gl}(4,\mathbb{R})$ valued 1-form: $\omega \in \Omega^1(F\mathcal{M},\mathfrak{gl}(4,\mathbb{R}))$.

Given some open subset $U\subseteq \mathcal{M}$ and a local section $s \colon U\to F\mathcal{M}$ (which is also referred to as \textit{local gauge}) we define the corresponding local connection 1-form $\omega_s \in \Omega^1(U,\mathfrak{gl}(4,\mathbb{R}))$ (which is the connection we usually work with on the base manifold) as the pullback
\begin{equation}
    \omega_s \coloneqq s^*\omega.
\end{equation}
The local connection 1-form can be expressed as
\begin{equation}
    \omega_s = \omega\indices{^a_b}\otimes E\indices{_a^b},
\end{equation}
where $\omega\indices{^a_b} \in \Omega^1(U)$ are called \textit{gauge boson fields} and $\{E\indices{_a^b}\}$ is a basis of $\mathfrak{gl}(4,\mathbb{R})$.
In a coordinate frame $\{\partial_\mu\}$ around $p\in U$ we set 
\begin{equation}
    \omega_{\mu}(x) \coloneqq \omega_s (p)(\partial_\mu) = \omega\indices{^a_b}(p)(\partial_\mu) E\indices{_a^b},
\end{equation}
where $x$ denotes the coordinates of $p$.
Thus
\begin{equation}
    \omega_{\mu}(x) = \omega\indices{^a_{b\mu}}(x) E\indices{_a^b},
\end{equation}
where $\omega\indices{^a_{b\mu}} (x)\coloneqq \omega\indices{^a_b}(p)(\partial_\mu)$ are smooth real valued functions on the coordinate patch corresponding to $U$. These are precisely the connection components that we usually work with.

For any principal bundle, one can construct associated vector bundles.
\begin{figure}
    \centering
    \begin{tikzpicture}[
        node distance=5cm,
        every node/.style={font=\small},
        box/.style={draw, rounded corners, thick, minimum height=2em, minimum width=3cm, align=center}
        ]

    \node[box] (liegroup) {Lie Groups\\(gauge groups)};
    \node[box, right of=liegroup] (rep) {Representations\\on Vector Spaces};
    \node[box, below=1cm of liegroup] (principal) {Principal Bundles};
    \node[box, right of=principal] (assoc) {Associated Vector Bundles\\(matter fields)};
    \node[box, below=1cm of principal] (conn) {Connections\\(gauge fields)};
    \node[box, right of=conn] (covder) {Covariant Derivatives\\(interaction/coupling)};
    
    \draw[-Stealth, thick] (liegroup) -- (rep);
    \draw[-Stealth, thick] (liegroup) -- (principal);
    \draw[-Stealth, thick] (principal) -- (assoc);
    \draw[-Stealth, thick] (conn) -- (covder);
    \draw[-Stealth, thick] (rep) -- (assoc);
    \draw[-Stealth, thick] (principal) -- (conn);
    \draw[-Stealth, thick] (assoc) -- (covder);

    \end{tikzpicture}
    \caption{Schematic relationship between gauge symmetry, principal bundles, and associated vector bundles (adapted from \cite{Hamilton2017}).
    Lie groups define the gauge symmetry and act via representations on vector spaces, giving rise to associated vector bundles that model matter fields. A principal bundle encodes the gauge symmetry geometrically, while a connection on the principal bundle represents the gauge field. This connection induces covariant derivatives on associated bundles, describing the coupling between gauge fields and matter fields.}
    \label{Princ_bund_Assoc_bund}
\end{figure}
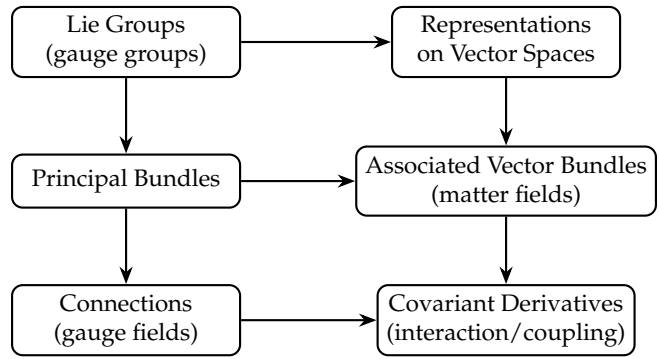
Generally, given any principal $G$-bundle $H$ and a representation $\rho\colon G \to \mathrm{Aut}(V)$ of $G$ on the vector space $V$, the associated vector bundle is defined as the quotient space
\begin{equation}
    (H \times V)/\sim,
\end{equation}
where the equivalence relation $\sim$ is given by
\begin{equation}
    (h,v) \sim (h\cdot g, \rho(g)^{-1}v)
\end{equation}
for any $h\in H$, $g\in G$, and $v\in V$.
These vector bundles play a central role in gauge theories, as classical matter fields are described by sections of such associated bundles, see Fig.\,\ref{Princ_bund_Assoc_bund}.

\subsection{Associated Vector Bundles and Covariant Derivatives}

Depending on the type of geometric object under consideration, different associated vector bundles arise naturally from the frame bundle. All classical fields on spacetime—such as vector, covector, and tensor fields—are most naturally understood as sections of such associated bundles, with their transformation properties determined by representations of the structure group $\mathrm{GL}(4,\mathbb{R})$.

Introducing associated bundles allows us to formulate the action of the affine connection in a unified and representation-theoretic manner. In particular, a connection on the frame bundle induces covariant derivatives on all associated bundles, thereby providing a systematic definition of parallel transport and differentiation for arbitrary tensor fields. This construction will be essential in the following, as it underlies the definitions of torsion, curvature, and non-metricity, as well as the coupling of geometric degrees of freedom to matter fields in MAG.

\subsubsection{Vector Fields}
Vector fields on spacetime arise as sections of the tangent bundle, 
\begin{equation}
    T\mathcal{M} = \bigsqcup_{p \in \mathcal{M}} T_p \mathcal{M},
\end{equation}
which itself admits a natural realization as an associated vector bundle to the frame bundle. In order to understand this realization,
consider again the standard representation of the general linear group
\begin{equation}
    \rho \colon \mathrm{GL}(4,\mathbb{R}) \to \mathrm{Aut}(\mathbb{R}^4),
\end{equation}
which we will denote simply by $\rho(g) = g$ for convenience.
Given the frame bundle $F\mathcal{M}$, the vector bundle associated to it via this representation is defined as
\begin{equation}
    (F\mathcal{M} \times \mathbb{R}^4)/\sim,
\end{equation}
where the equivalence relation is given by
\begin{equation}
    \big((p, f_p), v\big) \sim \big((p, f_p \circ g), g^{-1}v\big),
    \quad \text{for all } g \in \mathrm{GL}(4, \mathbb{R}).
\end{equation}
There exists a natural identification between the equivalence classes and tangent vectors via
\begin{equation}
    \big[((p, f_p), v)\big] \mapsto \big(p, f_p(v)\big) \in T\mathcal{M}.
\end{equation}
Thus, we obtain a vector bundle isomorphism
\begin{equation}
    T\mathcal{M} \cong (F\mathcal{M} \times \mathbb{R}^4)/\sim.
\end{equation}
In this way, the tangent bundle is naturally realized as an associated vector bundle to the frame bundle.

Now let $\tilde{V}$ be a section of the tangent bundle and $X$ be 
any vector field. Then we can write $\tilde{V} = [s,V]$, where $s \colon U\to F\mathcal{M}$ is some local gauge and $V\colon U \to \mathbb{R}^4$
is the component function of $\tilde{V}$ in the local gauge $s$.
We define the covariant derivative of $\tilde{V}$ wrt $\omega$, in the direction of $X$ as 
\begin{equation}
    \nabla_X^{\omega}\tilde{V} \coloneqq [s,\mathrm{d}V(X) + \rho_{\text{ind}}(\omega_s(X))V].
\end{equation}
This definition is independent of the choice of local gauge $s$.
Here, $\rho_{\text{ind}}\colon \mathfrak{gl}(4,\mathbb{R})\to \mathrm{End}(\mathbb{R}^4)$ is the induced Lie algebra representation of $\rho$, while $\mathrm{d}$ denotes the exterior derivative. Since $\rho$ is the standard representation in this case,
the action of $\rho_{\text{ind}}$ reduces to $\rho_{\text{ind}}(\omega_s(X))V = \omega_s(X)V$, which is to be understood as simple matrix-vector multiplication.
Choosing $X = \partial_\mu$ then yields the well-known form
\begin{equation}
    \mathrm{d}V(X) + \rho_{\text{ind}}(\omega_s(X))V = \partial_\mu V + \omega_{\mu}V,
\end{equation}
or 
\begin{equation}
     \partial_\mu V^a + \omega\indices{^a_{b\mu}}V^b
\end{equation}
in component notation.

\subsubsection{Covector Fields}
For covector fields the relevant associated vector bundle is the cotangent bundle $T^*\mathcal{M}$ whose total space is the disjoint union of cotangent spaces at every point on the manifold. In this case the relevant representation is 
the dual representation $\rho^* \colon \mathrm{GL}(4,\mathbb{R}) \to \mathrm{Aut}((\mathbb{R}^4)^*)$ of $\rho$. The dual of the standard representation is
\begin{equation}
    \rho^*(g) = (g^{-1})^\intercal.
\end{equation}
Its induced Lie algebra representation $\rho^*_{\text{ind}} \colon \mathfrak{gl}(4,\mathbb{R}) \to \mathrm{End}((\mathbb{R}^4)^*)$ is given by
\begin{equation}
    \rho^*_{\text{ind}}(A) = -A^\intercal.
\end{equation}
The minus sign ensures invariance of the natural pairing between vectors and covectors.

Now given a section of the cotangent bundle $\tilde{\alpha} = [s,\alpha]$
for some local gauge $s$ and its coordinate function $\alpha$, we define
the covariant derivative of $\tilde{\alpha}$ in the direction of a vector field $X$ as
\begin{equation}
    \nabla_X^{\omega}\tilde{\alpha} \coloneqq [s,\mathrm{d}\alpha(X) + \rho^*_{\text{ind}}(\omega_s(X))\alpha].
\end{equation}
Hence we recover the usual coordinate form with $X = \partial_\mu$ as
\begin{equation}
    \partial_\mu\alpha_a - \omega\indices{_a^{b}_\mu}\alpha_b.
\end{equation}

\subsubsection{Tensor Fields}
A $(k,l)$ tensor field is a section of the tensor bundle
\begin{equation}
    T^{(k,l)}\mathcal{M} = \bigotimes^k T\mathcal{M} \otimes \bigotimes^l T^*\mathcal{M}.
\end{equation}
The relevant representation is 
here given by a respective tensor product of representations of individual standard and dual representations
\begin{equation}
    \rho^{(k,l)} = \rho^{\otimes k} \otimes {(\rho^*)}^{\otimes l}.
\end{equation}
The induced Lie algebra representation acts independently on each tensor slot
\begin{equation}
    \rho^{(k,l)}_{\text{ind}} = \left(\sum_{i=1}^{k} \mathrm{id}^{\otimes(i-1)} \otimes \rho_{\text{ind}} \otimes \mathrm{id}^{\otimes(k - i)}\right) \otimes \mathrm{id}^{\otimes l}
    + \mathrm{id}^{\otimes k}\otimes \left(\sum_{j=1}^{l} \mathrm{id}^{\otimes(j-1)} \otimes \rho^*_{\text{ind}} \otimes \mathrm{id}^{\otimes(l - j)}\right).
\end{equation}
Thus each vector index transforms as a vector field and each covector index transforms as a covector field.
Take a section of the $(k,l)$ tensor bundle $\tilde{T} = [s,T]$
for some local gauge $s$ and its coordinate function $T$.
The covariant derivative of $\tilde{T}$ in the direction of the vector field $X$ is defined as
\begin{equation}
    \nabla_X^{\omega}\tilde{T} \coloneqq [s,\mathrm{d}T(X) + \rho^{(k,l)}_{\text{ind}}(\omega_s(X))T].
    \label{tens_cov_der}
\end{equation}
For $X = \partial_\mu$ we thus recover the well-known covariant derivative of tensor fields
\begin{equation}
    \partial_\mu T\indices{^{a_1\ldots a_k}_{b_1\ldots b_l}} + 
    \sum_{i = 1}^{k}\omega\indices{^{a_i}_{c\mu}}T\indices{^{a_1\ldots a_{i-1}\, c\, a_{i+1}\ldots a_k}_{b_1\ldots b_l}}
    - \sum_{j = 1}^{l}\omega\indices{_{b_j}^{c}_\mu}T\indices{^{a_1\ldots a_k}_{b_1\ldots b_{j-1}\, c \, b_{j+1}\ldots b_l}}.
\end{equation}

\subsection{Exterior Covariant Derivative}
While $\nabla^\omega$ acts naturally on sections, in order to define the concepts of curvature and torsion, we have to extend $\nabla^\omega$ to an exterior covariant derivative acting on bundle-valued forms.
Indeed, the previously defined $\nabla^\omega$ can be understood as a map $\nabla^\omega\colon \Gamma(E) \to \Omega^1(\mathcal{M},E)$ where $E$ denotes the total space of the associated bundle. More specifically, given a section $\sigma \in \Gamma(E)$,
$\nabla^{\omega}\sigma$ is an object that acts on a vector field and produces another section.
Explicitly,
\begin{equation}
    (\nabla^{\omega}\sigma)(X) \coloneqq \nabla^{\omega}_{X}\sigma \in \Gamma(E),
\qquad X \in \Gamma(T\mathcal{M}).
\end{equation}
As such we can think of $\nabla^\omega$ as a generalization of the standard exterior derivative $\mathrm{d}\colon C^\infty(\mathcal{M}) \to \Omega^1(\mathcal{M})$. Recall that it is possible to extend the exterior derivative to a map $\mathrm{d}\colon \Omega^k(\mathcal{M}) \to \Omega^{k+1}(\mathcal{M})$. In complete analogy, it is possible to extend the covariant derivative $\nabla^\omega$ to the \textit{exterior covariant derivative}
\begin{equation}
    \mathrm{d}_\omega\colon \Omega^k(\mathcal{M},E) \to \Omega^{k+1}(\mathcal{M},E).
\end{equation}

The procedure is as follows: we first define an exterior covariant derivative $\mathrm{D}_\omega$ on the principal bundle $P$. This in turn then determines 
$\mathrm{d}_\omega$ on the associated bundle via a canonical isomorphism. So consider a principal bundle $(P,\pi,\mathcal{M};G)$ with connection 1-form $\omega$
and a representation $\rho\colon G\to \mathrm{GL}(V)$ on some vector space $V$. Define the exterior covariant derivative as linear map  
\begin{equation}
    \mathrm{D}_\omega \colon \Omega^k(P,V) \to \Omega^{k+1}(P,V)
\end{equation}
with
\begin{equation}
(\mathrm{D}_\omega\alpha)_p(X_0,\ldots,X_{k}) \coloneqq \mathrm{d}\alpha_p (\mathrm{pr}_hX_0,\ldots,\mathrm{pr}_hX_k),
    \label{def_D_om}
\end{equation}
where $\mathrm{pr}_h$ denotes the projection onto the horizontal subspace defined by the connection~$\omega$.
Here $p\in P$, $X_i \in T_p P$ and $\mathrm{d}$ denotes the standard exterior derivative on $P$. 
A few remarks are in order. Since the principal bundle locally looks like $P \sim \mathcal{M}\times G$, the tangent space $T_p P$ at $p\in P$ contains directions for two kinds of motion:
\begin{itemize}
    \item \textbf{Vertical directions} – move along the fibre, i.e., change the gauge orientation but stay over the same base point $m\in\mathcal{M}$;
    \item \textbf{Horizontal directions} – move along the base manifold, i.e., change $m$ while keeping the gauge fixed.
\end{itemize}
The \emph{vertical subspace} $V_p P$ is the set of tangent vectors that do not change $m$:
\begin{equation}
    V_p P = \ker(\pi_{*p}).
\end{equation}
These correspond to infinitesimal gauge transformations.
A connection on $P$ tells us what it means to move ``horizontally'', that is, how to lift motion from the base manifold $\mathcal{M}$ into $P$ in a smooth and consistent way.
It gives a complementary \emph{horizontal subspace} $H_p P$ such that
\begin{equation}
    T_p P = H_p P \oplus V_p P.
\end{equation}
Any tangent vector $X \in T_p P$ thus splits uniquely as
\begin{equation}
    X = X_h + X_v,
\end{equation}
where $X_h \in H_p P$ is the horizontal component and $X_v \in V_p P$ is the vertical component.

The operator $\mathrm{pr}_h$ used in \eqref{def_D_om} projects the $X_i$ onto their horizontal components. One can then show that for any $\alpha \in \Omega_{hor}^k(P,V)^{(G,\rho)}$:
\begin{equation}
    \mathrm{D}_\omega\alpha = \mathrm{d}\alpha + \rho_{\text{ind}}(\omega)\wedge \alpha.
    \label{D_hor_equiv}
\end{equation}
The set $\Omega_{hor}^k(P,V)^{(G,\rho)}$ is the subset of elements in $\Omega^k(P,V)$ which vanish on vertical vectors and are $(G,\rho)$-equivariant.
Such forms are precisely the ones which can be ``projected down'' to the associated bundle. The wedge product appearing above is the usual wedge of differential forms, combined with the natural action of $\mathrm{End}(V)$ on $V$.
The expression \eqref{D_hor_equiv} makes explicit how the connection modifies the ordinary exterior derivative by a gauge-covariant correction term.
Relevant for us is how $\mathrm{D}_\omega$ pulls back to the base manifold $\mathcal{M}$. It turns out that, given a local gauge $s\colon U \to P$, the expression above becomes
\begin{equation}
    (\mathrm{D}_\omega \alpha)_s = \mathrm{d} \alpha_s + \rho_{\text{ind}}(\omega_s) \wedge \alpha_s.
\end{equation}
This expression will be useful to define curvature and the torsion 2-form.

This exterior covariant derivative $\mathrm{D}_\omega$ on the principal bundle now defines a related exterior covariant derivative on the associated bundle in the following manner.
There is a canonical isomorphism
\begin{equation}
    \Omega^k(\mathcal{M},E) \;\cong\; \Omega^k_{\mathrm{hor}}(P,V)^{(G,\rho)},
\end{equation}
between $E$-valued $k$-forms on the base manifold (recall: $E$ is the total space of the associated bundle) and horizontal, $(G,\rho)$-equivariant $V$-valued $k$-forms on the principal bundle. 
Explicitly, for each $\alpha \in \Omega^k(\mathcal{M},E)$, there exists a unique 
$\overline{\alpha} \in \Omega^k_{\mathrm{hor}}(P,V)^{(G,\rho)}$.
We then define the \emph{exterior covariant derivative} on the associated bundle as the operator
\begin{align}
    \mathrm{d}_\omega \colon \Omega^k(\mathcal{M},E) &\to \Omega^{k+1}(\mathcal{M},E)\nonumber\\
    \alpha &\mapsto \mathrm{d}_\omega \alpha,
\end{align}
such that its lift to the principal bundle satisfies
\begin{equation}
    \overline{\mathrm{d}_\omega \alpha} \; = \; \mathrm{D}_\omega \overline{\alpha}.
\end{equation}
In words, the derivative $\mathrm{d}_\omega$ is defined so that the diagram
\begin{center}
\begin{tikzcd}[column sep=large, row sep=large]
\Omega^k(\mathcal{M},E) 
  \arrow[r, "\mathrm{d}_\omega"] 
  \arrow[d, "\cong" left] 
  & 
\Omega^{k+1}(\mathcal{M},E) 
  \arrow[d, "\cong" right] \\[2pt]
\Omega^k_{\mathrm{hor}}(P,V)^{(G,\rho)} 
  \arrow[r, "\mathrm{D}_\omega"] 
  &
\Omega^{k+1}_{\mathrm{hor}}(P,V)^{(G,\rho)}
\end{tikzcd}
\end{center}
commutes.
Indeed, for $k=0$ the exterior covariant derivative reduces to the previously defined covariant derivative, $\mathrm{d}_\omega = \nabla^\omega$.

\subsection{Curvature}
Let us now precisely state what the curvature, or field strength, of a connection is, which in MAG encodes the gravitational field associated with the affine connection.
The curvature 2-form $F\in \Omega^2(P,\mathfrak{g})$ of the connection $\omega$ is given by 
\begin{equation}
    F = \mathrm{D}_\omega \omega.
\end{equation}
Explicitly, one can show that
\begin{equation}
    F = \mathrm{d}\omega + \frac{1}{2}[\omega,\omega],
\end{equation}
which is known as the \textit{second Cartan structure equation}.
Here, $\mathrm{d}$ denotes the exterior derivative and $[\cdot,\cdot]$ the Lie bracket. Note that these notions are extensions of the usual exterior derivative and Lie bracket since they are acting on Lie-Algebra-valued forms. For completeness, we briefly recall how these operations are defined. Since $\Omega^k(P,\mathfrak{g})\cong \Omega^k(P) \otimes \mathfrak{g}$ we can express any $\alpha \in \Omega^k(P,\mathfrak{g})$ as
\begin{equation}
    \alpha = \sum_{a=1}^{\mathrm{dim}(\mathfrak{g})} \alpha^a\otimes T_a,
\end{equation}
where $\{T_a\}$ is a basis for $\mathfrak{g}$ and $\alpha^a\in \Omega^k(P)$ the associated
$k$-forms to $\alpha$. Then we define
\begin{equation}
    \mathrm{d}\alpha \coloneqq \sum_{a=1}^{\mathrm{dim}(\mathfrak{g})} \mathrm{d}\alpha^a\otimes T_a,
\end{equation}
with the usual exterior derivative on the rhs, 
and for any $\beta \in \Omega^l(P,\mathfrak{g})$
\begin{equation}
    [\alpha,\beta] \coloneqq \sum_{a,b=1}^{\mathrm{dim}(\mathfrak{g})} \alpha^a\wedge \beta^b \otimes [T_a,T_b],
\end{equation}
with the usual Lie bracket and wedge product on the rhs.

In our specific case of the frame bundle we have $\omega = \sum_{a,b} \omega\indices{^a_b}\otimes E\indices{_a^b}$ with $\omega\indices{^a_b}\in \Omega^1(U)$, where $U\subset \mathcal{M}$, and $\{E\indices{_a^b}\}$ a basis of $\mathfrak{gl}(4,\mathbb{R})$. Thus 
\begin{equation}
    \mathrm{d}\omega = \sum_{a,b}  \mathrm{d}\omega\indices{^a_b}\otimes E\indices{_a^b}
\end{equation}
and 
\begin{equation}
    [\omega,\omega] = \sum_{a,b,c,d} \omega\indices{^a_b}\wedge \omega\indices{^c_d} \otimes [E\indices{_a^b},E\indices{_c^d}].
\end{equation}
The Lie bracket of the $\mathfrak{gl}(4,\mathbb{R})$ generators is given by $[E\indices{_a^b},E\indices{_c^d}] = \delta^b_c E\indices{_a^d} - \delta^d_a E\indices{_c^b}$. After some algebra and re-labelling indices we find
\begin{equation}
    [\omega,\omega] = 2\sum_{a,b,d} (\omega\indices{^a_b}\wedge \omega\indices{^b_d}) \otimes E\indices{_a^d}.
\end{equation}
Hence the curvature 2-form is given by
\begin{equation}
    F = \sum_{a,b}\left(\mathrm{d}\omega\indices{^a_b} + \sum_{d}
    \omega\indices{^a_d}\wedge \omega\indices{^d_b}\right)\otimes 
    E\indices{_a^b} \in \Omega^2(F\mathcal{M},\mathfrak{gl}(4,\mathbb{R})),
\end{equation}
with corresponding real-valued 2-forms
\begin{equation}
    F\indices{^a_b} = \mathrm{d}\omega\indices{^a_b} + \sum_{d}
    \omega\indices{^a_d}\wedge \omega\indices{^d_b} \in \Omega^2(U).
\end{equation}
In the context of gravity, the field strength of the affine connection corresponds to the spacetime curvature and generalizes the Riemann curvature of Levi-Civita geometry. It is
usually denoted by $R$. We shall adopt this notation henceforth.

\subsection{Soldering}
To define torsion in the metric-affine framework, one further geometric ingredient is required: the \textit{solder form}.
It provides the fundamental link between the principal bundle description of the theory and the geometry of spacetime.

The frame bundle carries a \textit{canonical 1-form} $\theta$ with 
values in $\mathbb{R}^4$, i.e., $\theta \in \Omega^1(F\mathcal{M},\mathbb{R}^4)$. Given any $u\in F\mathcal{M}$, it is defined by
\begin{equation}
    \theta_u(X)\coloneqq u^{-1}(\pi_*(X)),
\end{equation}
for any $X \in T_u F\mathcal{M}$. 
Recall that any point in $F\mathcal{M}$ consists of a pair which includes a point $p\in \mathcal{M}$ and a vector space isomorphism $\mathbb{R}^4 \to T_p \mathcal{M}$. With $u^{-1}$ we denote here the inverse of this isomorphism, i.e., a map $T_p \mathcal{M}\to \mathbb{R}^4$.
Furthermore, $\pi$ is the projection of the frame bundle and $\pi_*$ its pushforward. Intuitively, $\theta$ provides the components of the vector $X$ with respect to the frame $u$. Such a $\theta$ is more generally called a \textit{solder form}. A solder form glues the fibres of the principal bundle to the base manifold. In this sense it relates the tangent spaces of the base manifold to the fibres. Since here we are working with the frame bundle, the fibres (ordered bases of the tangent spaces) are related to the tangent spaces very naturally.

Given some gauge $\sigma\colon \mathcal{M}\to F\mathcal{M}$, the pullback
\begin{equation}
    \sigma^* \theta \in \Omega^1(\mathcal{M})
\end{equation}
turns out to be the coframe. Indeed, for any $p\in \mathcal{M}$
and $v\in T_p\mathcal{M}$ we find 
\begin{equation}
    (\sigma^* \theta)_p(v) = \sigma(p)^{-1}(v).
\end{equation}
Since $\sigma(p)\in F\mathcal{M}$, we have that $\sigma(p)^{-1}\colon T_p\mathcal{M}\to \mathbb{R}^4$. In fact 
$\sigma(p)^{-1}(v)$ just gives the components of $v$ wrt the frame chosen by $\sigma$ at $p$. Thus we understand $(\sigma^* \theta)_p^a \in T^*_p\mathcal{M}$ as providing the $a$-th component of $v$.
Hence as a map we have 
\begin{equation}
    (\sigma^* \theta)^a = e^a,
\end{equation}
where $e^a$ is the coframe (not necessarily a tetrad).

The solder form provides a natural geometric definition of torsion as we will see in the following.

\subsection{Curvature, Torsion, and Non-metricity}
In MAG, the fundamental geometric variables are the Lorentzian metric $g$, an independent linear connection $\omega\indices{^a_b}$, and the coframe $e^a$
(the local representative of the solder form $\theta$). The metric $g$ is a section of the symmetric $(0,2)$ tensor bundle and
defines lengths and angles on spacetime. The
connection $\omega\indices{^a_b}
= \omega\indices{^a_{b\mu}}\,\mathrm{d}x^\mu$ is a set of 1-forms and defines parallel transport and
the affine structure independently of the metric. Finally, the coframe
$e^a = e\indices{^a_\mu}\,\mathrm{d}x^\mu$ is also a set of
$1$-forms which however defines a local frame and implements the soldering between
the tangent bundle of spacetime and the internal geometry, here given by $\mathbb{R}^4$ with structure group $\mathrm{GL}(4,\mathbb{R})$. 
These three objects allow us to define curvature, torsion, and non-metricity in a very natural manner:
\begin{align*}
  \text{Curvature}     &\colon \hspace{-2cm} & R &\coloneqq \mathrm{D}_\omega \omega = \mathrm{d}\omega + \frac{1}{2}[\omega,\omega]\\
  \text{Torsion}       &\colon \hspace{-2cm} & S &\coloneqq \mathrm{D}_\omega \theta = \mathrm{d}\theta + \omega \wedge \theta\\
  \text{Non-metricity}  &\colon \hspace{-2cm} & Q &\coloneqq \mathrm{d}_\omega g = \mathrm{d} g + \rho^{(0,2)}_{\text{ind}}(\omega)g
\end{align*}
Here $R$ and $S$ are $\mathfrak{gl}(4)$- and $\mathbb{R}^4$-valued
differential forms on the frame bundle, while non-metricity $Q$ is
defined as a tensorial object on the associated tensor bundle. The solder
form $\theta$ can be shown to be horizontal and
$\mathrm{GL}(4,\mathbb{R})$-equivariant; hence the application of
$\mathrm{D}_\omega$ reduces to the standard formula for horizontal
equivariant forms, cf.~\eqref{D_hor_equiv}. In this case, the induced
representation $\rho_{\mathrm{ind}}$ coincides with the fundamental
representation, which is why it is omitted explicitly. Furthermore, since
the metric is a $k=0$ form, one has $\mathrm{D}_\omega g = \nabla^{\omega}g$,
and \eqref{tens_cov_der} may be used, which coincides with the usual
covariant derivative of the metric.

In order to recover the familiar tensorial expressions on the base
manifold, the geometric objects above must be
pulled back using local sections. Let $\sigma\colon U\to F\mathcal{M}$ be
a local gauge. The corresponding spacetime expressions are then
obtained by applying the pullback $\sigma^*$.
For torsion, the
pullback yields
\begin{equation}
  S^a \equiv \sigma^* S^a
  = \mathrm{d}e^a + \omega\indices{^a_b}\wedge e^b .
\end{equation}
In local coordinates this becomes
\begin{equation}
  S^a
  = 2\Bigl(
      \partial_{[\mu} e\indices{^a_{\nu]}}
      + \omega\indices{^a_{b[\mu}}\,e\indices{^b_{\nu]}}
    \Bigr)
    \mathrm{d}x^\mu \otimes\mathrm{d}x^\nu ,
\end{equation}
from which we identify the torsion components
\begin{equation}
  S\indices{^a_{\mu\nu}}
  = 2\Bigl(
      \partial_{[\mu} e\indices{^a_{\nu]}}
      + \omega\indices{^a_{b[\mu}}\,e\indices{^b_{\nu]}}
    \Bigr).
\end{equation}
Using the relation 
\begin{equation}
  	\omega\indices{^a_{b\mu}}
  = e\indices{^a_\rho}\,\partial_\mu e\indices{_b^\rho} + e\indices{^a_\rho}\,\Gamma\indices{^\rho_{\nu\mu}}\,e\indices{_b^\nu},
\end{equation}
where $\Gamma$ denotes the connection in local coordinates,
we obtain the spacetime torsion tensor
\begin{equation}
  S\indices{^\lambda_{\mu\nu}}
  \equiv e\indices{^\lambda_a} S\indices{^a_{\mu\nu}}
  = 2\,\Gamma\indices{^\lambda_{[\mu\nu]}} .
\end{equation}
In the following, we however adopt the commonly used convention
\begin{equation}
  S\indices{^\lambda_{\mu\nu}}
  \coloneqq \Gamma\indices{^\lambda_{[\mu\nu]}},
\end{equation}
by absorbing the factor two into the definition of the torsion tensor.
The curvature two-form pulls back as
\begin{equation}
  R\indices{^a_b}
  = \mathrm{d}\omega\indices{^a_b}
    + \omega\indices{^a_d}\wedge\omega\indices{^d_b},
\end{equation}
which yields the familiar coordinate expression
\begin{equation}
  R\indices{^\rho_{\sigma\mu\nu}}
  = \partial_\mu \Gamma\indices{^\rho_{\sigma\nu}}
    - \partial_\nu \Gamma\indices{^\rho_{\sigma\mu}}
    + \Gamma\indices{^\rho_{\lambda\mu}}\,\Gamma\indices{^\lambda_{\sigma\nu}}
    - \Gamma\indices{^\rho_{\lambda\nu}}\,\Gamma\indices{^\lambda_{\sigma\mu}} .
\end{equation}
Lastly, the non-metricity tensor is given by the covariant derivative of the
metric,
\begin{equation}
  Q\indices{_{\mu\nu\rho}}
  = \partial_\mu g\indices{_{\nu\rho}}
    - \Gamma\indices{^\sigma_{\nu\mu}}\,g\indices{_{\sigma\rho}}
    - \Gamma\indices{^\sigma_{\rho\mu}}\,g\indices{_{\nu\sigma}}
  = \nabla_\mu g\indices{_{\nu\rho}} .
\end{equation}

Together, torsion and non-metricity allow for the general decomposition of
the affine connection,
\begin{equation}
  \Gamma\indices{^\lambda_{\mu\nu}}
  = \mathring{\Gamma}\indices{^\lambda_{\mu\nu}}
    + K\indices{^\lambda_{\mu\nu}}
    + L\indices{^\lambda_{\mu\nu}},
\end{equation}
where $\mathring{\Gamma}\indices{^\lambda_{\mu\nu}}$ is the Levi--Civita
connection,
\begin{equation}
  \mathring{\Gamma}\indices{^\lambda_{\mu\nu}}
  = \tfrac{1}{2} g^{\lambda\sigma}
    \bigl(
      \partial_\mu g\indices{_{\nu\sigma}}
      + \partial_\nu g\indices{_{\mu\sigma}}
      - \partial_\sigma g\indices{_{\mu\nu}}
    \bigr),
\end{equation}
$K\indices{^\lambda_{\mu\nu}}$ is the contortion tensor,
\begin{equation}
  K\indices{^\lambda_{\mu\nu}}
  = S\indices{^\lambda_{\mu\nu}}
    - S\indices{_\mu^{\lambda}_{\nu}}
    - S\indices{_\nu^{\lambda}_{\mu}},
\end{equation}
and $L\indices{^\lambda_{\mu\nu}}$ denotes the disformation,
\begin{equation}
  L\indices{^\lambda_{\mu\nu}}
  = \tfrac{1}{2} g^{\lambda\rho}
    \bigl(
      - Q\indices{_{\mu\rho\nu}}
      + Q\indices{_{\rho\nu\mu}}
      + Q\indices{_{\nu\rho\mu}}
    \bigr).
\end{equation}
It is worth noting that the affine connection in metric-affine geometry admits a projective gauge freedom which leave the autoparallel structure invariant and play an important role in several formulations of metric-affine and Weylian geometries \cite{Sauro22}.

For the majority of this review we focus on torsional effects and
therefore restrict attention to metric-compatible connections,
$Q\indices{_{\lambda\mu\nu}}=0$, for which the affine connection reduces
to
\begin{equation}
  \Gamma\indices{^\lambda_{\mu\nu}}
  = \mathring{\Gamma}\indices{^\lambda_{\mu\nu}}
    + K\indices{^\lambda_{\mu\nu}} .
\end{equation}
In the final part of this review, however, we will also encounter
connections that are torsion-free but not metric-compatible, as they
naturally arise in the context of information geometry.

This completes the geometric framework underlying MAG and fixes the notation used throughout the later parts of this review.

\section{Classical Variation and Optimal Control of Velocity Field} \label{sec:clas}

To set the stage for SVM, we first recast the familiar classical variational principle as an optimal control problem for a velocity field.
Let ${\bf r}(t)$ denote the trajectory of a particle in a $D$-dimensional Cartesian space. 
The Lagrangian for this single-particle system is
\begin{equation}
L = K - V =  \frac{\uM}{2} \left( \frac{\ud{\bf r}(t)}{\ud t} \right)^2 - V({\bf r}(t))\,, \label{eqn:cla-lag}
\end{equation}
where $\uM$ is the particle mass and $V$ denotes a potential energy.
For later convenience, 
the velocity field ${\bf u}({\bf x}, t)$ is defined such that it coincides with the particle's velocity along its trajectory:
\begin{equation}
\frac{\ud{\bf r}(t)}{\ud t} = {\bf u}({\bf r}(t), t)\,. \label{eqn:cla-vel}
\end{equation}

Hamilton's principle states that the action, which is defined by the time integral of the Lagrangian, is stationary with respect to variations of the trajectory. An infinitesimal variation of the trajectory is given by
\begin{equation}
{\bf r}(t) \to {\bf r}'(t) = {\bf r}(t) + \delta{\bf f}({\bf r}(t), t)\,, \label{eqn:cla-vari}
\end{equation}
where $\delta{\bf f}({\bf x}, t)$ is an infinitesimal smooth vector field representing the virtual displacement, vanishing at the boundaries: $\delta{\bf f}({\bf x}, t_i) = \delta{\bf f}({\bf x}, t_f) = 0$.
After performing an integration by parts, the variation of the action is found to be
\begin{equation}
\delta I[{\bf r}] = \int_{t_i}^{t_f} \ud t \, \left[ -\uM \left( \frac{\ud}{\ud t}{\bf u}({\bf r}(t), t) \right) - \nabla V({\bf r}(t)) \right] \cdot \delta{\bf f}({\bf r}(t), t)\,. \label{eqn:deltaI_cla}
\end{equation}
In this calculation, we used Eq.\ (\ref{eqn:cla-vel}) and $\ud/\ud t = \partial_t + {\bf u} ({\bf r}(t), t)\cdot \nabla$ being the material derivative at ${\bf r}(t)$.
Since the variation $\delta{\bf f}$ is arbitrary, the stationary condition $\delta I[{\bf r}]=0$ requires the term in the brackets to vanish for any point on the trajectory. This yields the equation for the velocity field:
\begin{equation}
(\partial_t  + {\bf u}({\bf x},t)\cdot \nabla ){\bf u} ({\bf x},t)= - \frac{1}{\uM}\nabla V({\bf x}) \, . 
\end{equation}
In the present case, the particle trajectory is deterministic and thus it is enough to define the velocity at the position of the particle. 
Using Eq.\ (\ref{eqn:cla-vel}) again and evaluating this field equation along the particle's trajectory ${\bf r}(t)$ gives Newton's second law:
\begin{equation}
\uM \frac{\ud^2 {\bf r}(t)}{\ud t^2} = - \nabla V({\bf r}(t))\, .
\end{equation}

\section{General Setup for Stochastic Variation} \label{sec:svm_general}

In the previous section, we implicitly assumed that the particle trajectory is smooth and hence differentiable.
As mentioned in the introduction, several proposals exist to formulate the variation of non-differentiable trajectories 
\cite{A_Holland_1977,O_Inoue_1979,O_Ioannis_1980,S_Yasue_1981,S_Nakagomi_1981,S_Papiez_1982,A_Yasue_1983,S_Fleming_1983,Q_Guerra_1983,A_Rosenbrock_1985,V_Marra_1987,R_Serva_88,F_KIME_89,L_Jaekel_1990,H_Pavon_1995,Nagasawa_book_00,A_Aguiar_2005,L_KAPPEN_05,S_Cresson_2007,S_Eyink_2010,L_Arnaudon_2012,N_Koide_2012,F_Delbaen_2015,V_Holm_2015,R_Novikov_2018,C_Marner_2019,N_Bela_2019,A_Ohsumi_2019,Q_Lindgren_2019,N_Koide_2019,V_Koide_2020,Armin_Koide_2025}.
In this work, we follow the method proposed by Yasue \cite{S_Yasue_1981}, who introduced this idea to extend the formulation of Nelson's stochastic mechanics \cite{D_Nelson_1966}.
See also the review papers \cite{S_Zambrini_1985,U_Koide_2015,S_Kodama_2022,U_Goncalves_2020}. 
For a review of selected topics on stochastic approaches to hydrodynamics, see Ref.\ \cite{S_Bela_2020}.
As a recent book on Nelson's stochastic mechanics, see \cite{Kuipers2023}.

Applying the variational principle to non-differentiable trajectories requires two essential generalizations: the introduction of distinct forward and backward stochastic processes (see Eqs.\ (\ref{eqn:fsde}) and (\ref{eqn:bsde})), and a corresponding generalization of the time derivative (see Eqs.\ (\ref{eqn:mfd}) and (\ref{eqn:mbd})).

\subsection{Stochastic Kinematics of Forward and Backward Processes}

\begin{figure}[t]
\begin{center}
%\vspace*{-3cm}
\includegraphics[scale=0.4]{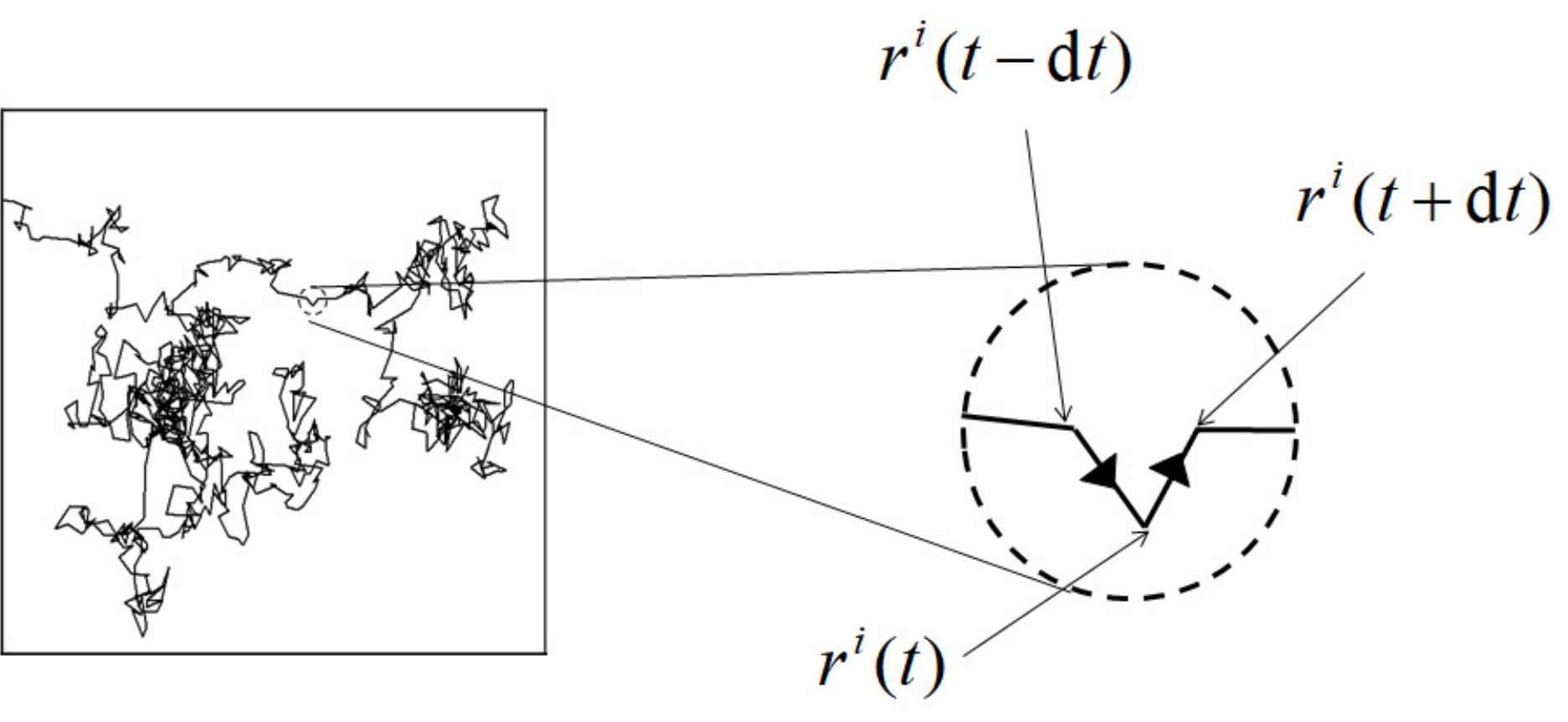}
\end{center}
\caption{An example of the typical trajectory of Brownian motion. The positions at $t- \ud t$, $t$ and $t+ \ud t$ are denoted by $r^{i}(t-\ud t)$, $r^{i}(t)$ and $r^{i}(t+\ud t)$\, , respectively.}
\label{fig:tra1}
\end{figure}

In contrast to the smooth, differentiable trajectories of classical mechanics, 
we now consider trajectories that are continuous but non-differentiable. 
A canonical example of such a path is the trajectory of a particle undergoing Brownian motion, as schematically depicted in Figure \ref{fig:tra1}. 

The most direct way to formalize this is to generalize the classical kinematic relation, 
Eq.~(\ref{eqn:cla-vel}), by incorporating a stochastic noise term. 
This leads to the forward stochastic differential equation (SDE) of the It\^o type, which describes the time evolution of the trajectory:
\begin{eqnarray}
\ud \widetilde{\bf r}(t) =  {\bf u}_+ (\widetilde{\bf r}( t), t) \ud t + \sqrt{2\nu} \ud\widetilde{\bf B}_+ (t)  \, \, \, (\ud t > 0) \, . \label{eqn:fsde}
\end{eqnarray} 
Here, the hat symbols like $\widetilde{\bf r}(t)$ denote stochastic quantities, and infinitesimal displacements are defined for an arbitrary 
stochastic quantity $\widetilde{A}(t)$ as 
$\ud \widetilde{A}(t) \equiv \widetilde{A}(t+ \ud t) - \widetilde{A}(t)$.
The smooth vector field ${\bf u}_+({\bf x},t)$, which will be determined by the stochastic variation, should not be confused with the instantaneous velocity of the particle, which is ill-defined for a non-differentiable trajectory. 
The time is discretized with the step $\ud t$, but we should take the limit $\ud t \to 0+$ at the end of calculations.
The second term on the right-hand side represents the Wiener process $\widetilde{\bf B}_+(t)$, which is the origin of the zigzag motion. The standard Wiener process is characterized by the following correlation properties for discretized time steps:
\begin{equation}
\begin{split}
\uE[ \ud\widetilde{\bf B}_+(t) ] &= 0\, , \\
\uE [ \ud\widetilde{B}^{i}_+ (t) \ud\widetilde{B}^{j}_+ (t^\prime) ] &= \ud t \, \delta_{t,t^\prime} \, \delta_{i,j}  \, , 
\end{split}
\label{eqn:f_wiener}
\end{equation}
where $\uE[\cdot]$ denotes the ensemble average. 
The intensity of the fluctuations is controlled by a non-negative constant $\nu$. 
For a detailed description of the Wiener process, see, e.g., Ref.\ \cite{Gardiner_2004}. 
Note that while the field ${\bf u}_+ ({\bf x},t)$ is a smooth function, its value along the trajectory, ${\bf u}_+ (\widetilde{\bf r}(t), t)$, is a stochastic variable.

\begin{figure}[t]
\begin{center}
%\vspace*{-3cm}
\includegraphics[scale=0.3]{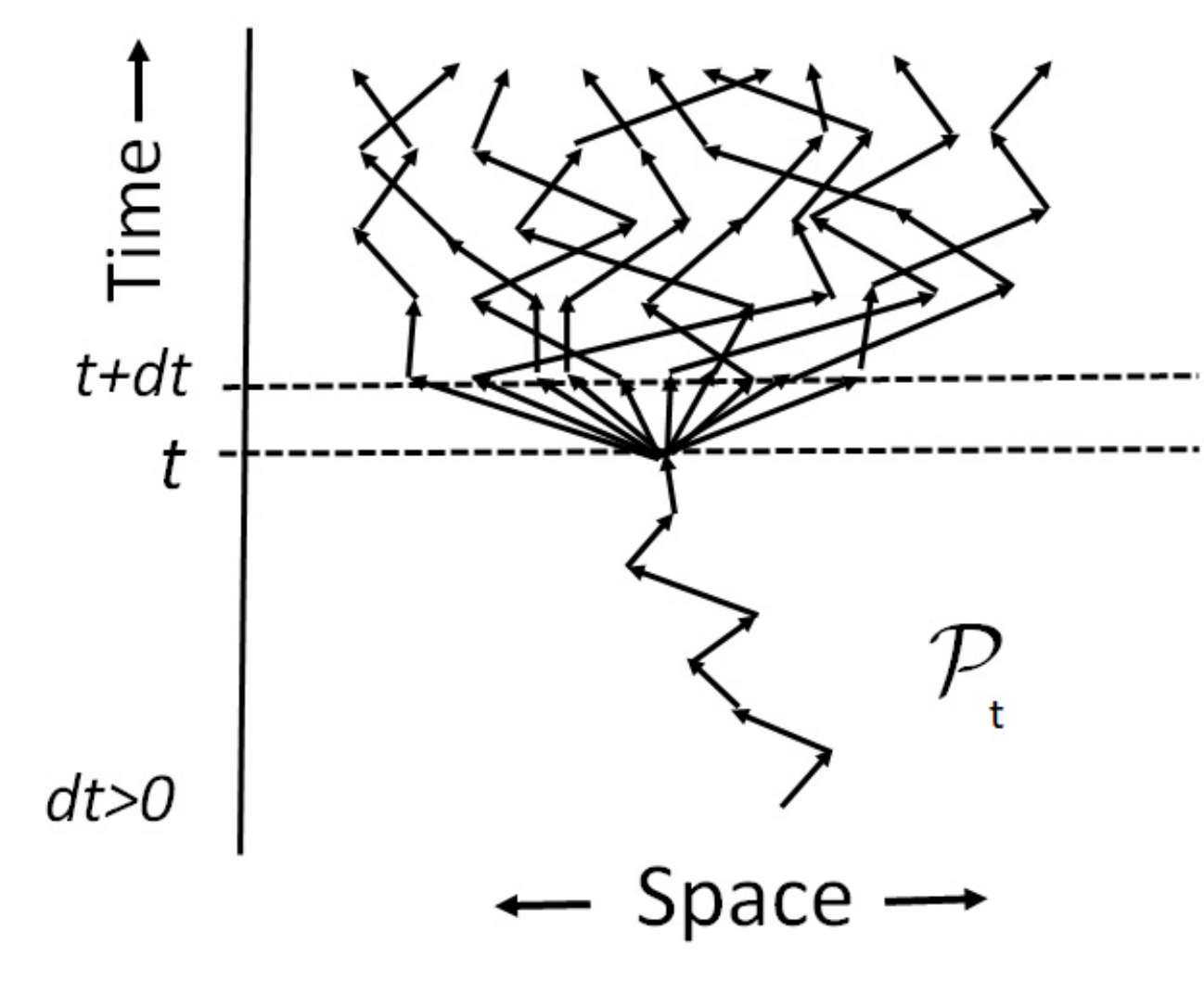}
\includegraphics[scale=0.3]{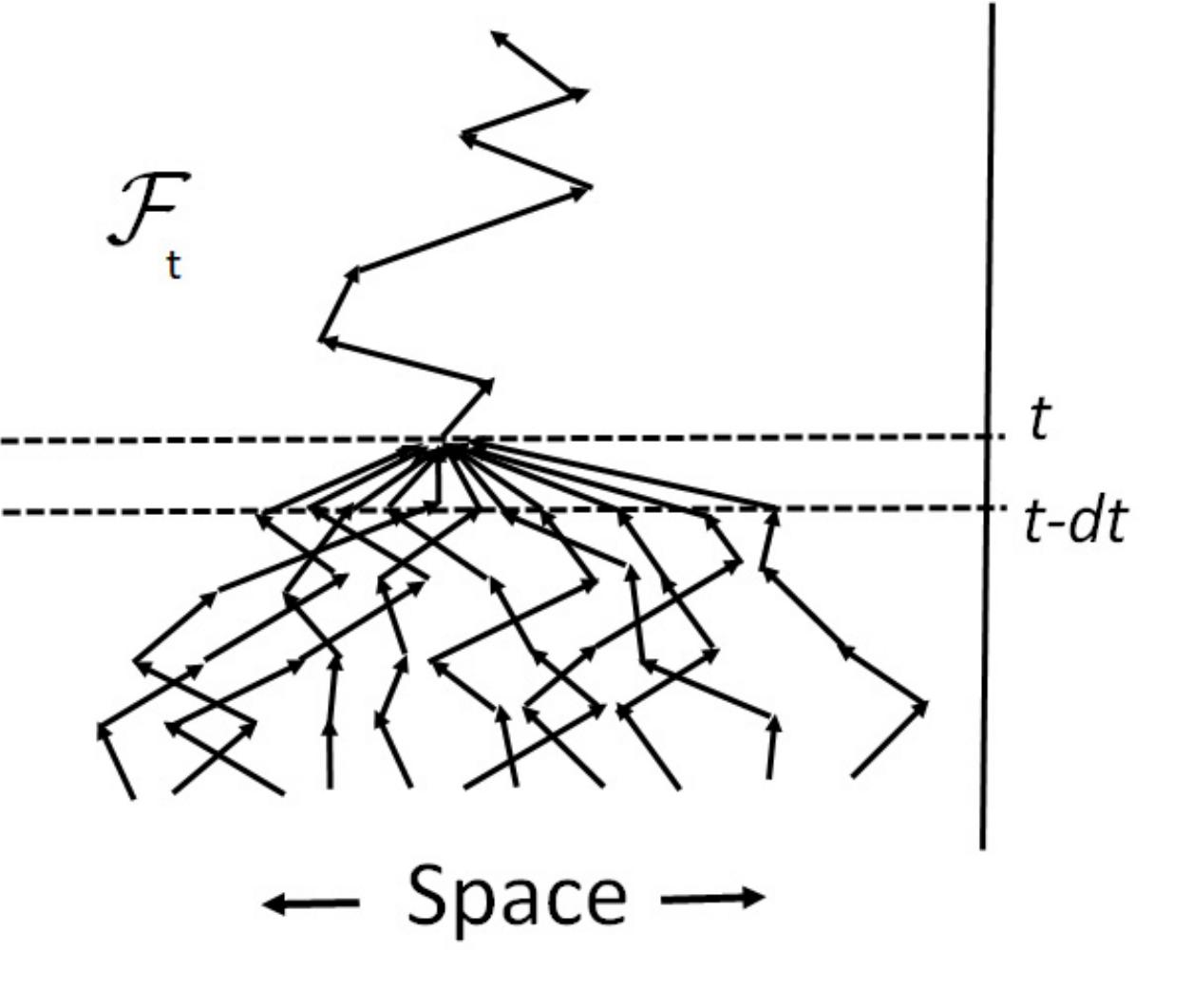}
\end{center}
\caption{The ensembles of trajectories fixing ${\cal P}_t$ and ${\cal F}_t$ shown in 
the left and right panels, respectively. 
This figure is taken from Fig. 2 of Ref. \cite{U_Goncalves_2020}.}
\label{fig:tra2}
\end{figure}

A crucial consequence of non-differentiability is that the instantaneous velocity 
is not uniquely defined. 
As illustrated in Fig.\ \ref{fig:tra1}, one can define a ``forward velocity" by observing the particle's displacement into the immediate future, ${\displaystyle \lim_{\ud t \rightarrow 0+}}(\widetilde{\bf r}(t+\ud t) - \widetilde{\bf r}( t))/\ud t$, 
or a ``backward velocity" by observing its displacement from the immediate past, ${\displaystyle \lim_{\ud t \rightarrow 0+}}(\widetilde{\bf r}(t) - \widetilde{\bf r}(t-\ud t))/\ud t$. 
This leads to the introduction of two distinct time derivatives \cite{D_Nelson_1966}. 
The mean forward derivative $D_+$ is defined as the expectation conditioned on the past trajectory, or more formally, on the filtration $\mathcal{P}_t = \{\widetilde{\bf r}(s) | s \le t \}$:
\begin{equation}
\uD_+  f  ( \widetilde{\bf r}( t) )  = \lim_{\ud t \rightarrow0+} \uE \left[  \frac{ f(\widetilde{\bf r}({t + \ud t})) -
f(\widetilde{\bf r}(t))}{\ud t} \Big| \mathcal{P}_{t} \right]\,  , \label{eqn:mfd}
\end{equation}
To be more precise, the conditioning is on the $\sigma$-algebra $\mathcal{P}_{t}$ generated by the process up to time t. This $\mathcal{P}_{t}$ forms an increasing family of sub-$\sigma$-algebras, known as a filtration.
For a Markov process, this conditioning simplifies to fixing the present position, $\mathcal{P}_t \to \widetilde{\bf r}(t)$. 
Applying this operator to $\widetilde{\bf r}(t)$, and using the SDE (\ref{eqn:fsde}), the mean forward velocity is 
\begin{equation}
\uD_+  \widetilde{\bf r}(t)  = \lim_{\ud t \rightarrow0+} \uE \left[  {\bf u}_+ (\widetilde{\bf r}(t),t) + \sqrt{2\nu} \frac{\ud \widetilde{\bf B}_+(t)}{\ud t} \Big| \widetilde{\bf r}(t) \right] =  {\bf u}_+ (\widetilde{\bf r}(t),t) \, . \label{eqn:fvel_con}
\end{equation}
Figure \ref{fig:tra2} illustrates the physical meaning of the mean derivatives.
We further introduce the mean backward derivative $D_-$ associated with the ``backward velocity", which is conditioned on the future trajectory $\mathcal{F}_t = \{\widetilde{\bf r}(s) | s \ge t \}$:
\begin{equation}
\uD_-  f(\widetilde{\bf r}(t))  = \lim_{\ud t \rightarrow0-} \uE \left[  \frac{ f(\widetilde{\bf r}({t + \ud t})) -
f(\widetilde{\bf r}(t))}{\ud t} \Big| \mathcal{F}_{t} \right]\, , \label{eqn:mbd}
\end{equation}
where $\mathcal{F}_{t}$ forms a decreasing family of sub-$\sigma$-algebra.
This is however not applicable to Eq.\ (\ref{eqn:fsde}) because $\ud t >0$ there.
To calculate the above derivative, we have to introduce a backward SDE, which describes the stochastic process equivalent to Eq.\ (\ref{eqn:fsde}) but describes the backward time evolution $\ud t<0$.
Suppose that the backward SDE is given by 
\begin{eqnarray}
\ud \widetilde{\bf r}(t) =  {\bf u}_- (\widetilde{\bf r}( t), t) \ud t + \sqrt{2\nu} \ud\widetilde{\bf B}_- (t)  \, \, \, (\ud t < 0) \, , \label{eqn:bsde}
\end{eqnarray} 
where
\begin{equation}
\begin{split}
\uE[ \ud\widetilde{\bf B}_-(t) ] &= 0\, , \\
\uE [ \ud\widetilde{B}^{i}_- (t) \ud\widetilde{B}^{j}_- (t^\prime) ] &= |\ud t| \, \delta_{t,t^\prime} \, \delta_{i,j}   \, , 
\end{split}
\label{eqn:b_wiener}
\end{equation}
The absolute value in the second correlation is due to $\ud t <0$.
Again, for a Markov process, the condition simplifies to $\mathcal{F}_t \to \widetilde{\bf r}(t)$ and then ${\bf u}_- (\widetilde{\bf r}(t),t)$ is the mean backward velocity:
\begin{equation}
\uD_-  \widetilde{\bf r}(t)  = \lim_{\ud t \rightarrow0-} \uE \left[  {\bf u}_- (\widetilde{\bf r}(t),t) + \sqrt{2\nu} \frac{\ud \widetilde{\bf B}_- (t)}{\ud t} \Big| \widetilde{\bf r}( t) \right] =  {\bf u}_- (\widetilde{\bf r}(t),t)\, . \label{eqn:bvel_con}    
\end{equation}
This dual description using forward and backward processes is a cornerstone of this formalism. The two velocities are not independent but are related through the probability density $\rho({\bf x},t)$ of the process, as seen soon later.

A key tool for the variational principle is the stochastic integration by parts formula, which relates the two derivatives:
\begin{align}
\int^{t_f}_{t_i} \ud t  \, 
\uE\left[
\widetilde{Y}(t) \uD_+ \widetilde{X}(t) + \widetilde{X}(t) \uD_- \widetilde{Y}(t)
\right] 
&= 
\lim_{n \rightarrow \infty} \sum_{j=0}^{n-1} \uE\left[
(\widetilde{X}(t_{j+1}) - \widetilde{X}(t_j)) \frac{\widetilde{Y}(t_{j+1}) + \widetilde{Y}(t_j)}{2}
\right] \nonumber \\
& + 
\lim_{n \rightarrow \infty} \sum_{j=1}^{n} \uE\left[
(\widetilde{Y}(t_{j}) - \widetilde{Y}(t_{j-1})) \frac{ \widetilde{X}(t_j) + \widetilde{X}(t_{j-1})}{2}
\right] \nonumber \\
&=
\lim_{n \rightarrow \infty} \sum_{j=0}^{n-1} \uE\left[
\widetilde{X}(t_{j+1}) \widetilde{Y}(t_{j+1})-\widetilde{X}(t_{j})\widetilde{Y}(t_{j})
\right]  \nonumber \\
&=
\uE\left[
\widetilde{X}(t_f)\widetilde{Y}(t_f) - \widetilde{X}(t_i)\widetilde{Y}(t_i)
\right] \, , \label{eqn:spif}
\end{align}
where $t_j = t_i + j \frac{t_f - t_i}{n} = t_i + j \ud t$.

\subsection{Fokker-Planck Equation and Consistency Condition} \label{sec:fp}

For the forward and backward SDEs to describe the same underlying stochastic process, the fields ${\bf u}_\pm ({\bf x},t)$ cannot be independent. 
This consistency requirement is formulated by examining the dynamics of the probability density $\rho(x,t)$:
\begin{eqnarray}
\rho ({\bf x},t) = \int \ud^D {\bf R} \,  \rho_0 ({\bf R}) \uE\left[
\delta^{(D)} ({\bf x} - \widetilde{\bf r}(t))
\right]   \, , \label{eqn:rho}
\end{eqnarray}
where ${\bf R} = \widetilde{r}(t_i)$ denotes the position of the Brownian particle at a given initial time $t_i$ and 
$\rho_0 ({\bf R})$ is the initial probability distribution satisfying 
\begin{eqnarray}
\int \ud^D {\bf R} \, \rho_0 ({\bf R}) = 1 \, .
\end{eqnarray}

As is well-known, the evolution equation of $\rho ({\bf x},t)$ is obtained by using the SDE's and It\^o's lemma \cite{Gardiner_2004} which is a truncated Taylor series expansion for a function of stochastic variables. 
For example, the evolution equation for the Dirac delta function is 
\begin{align}
& \ud \delta^{(D)} ({\bf x} - \widetilde{\bf r}(t))\nonumber \\
&= 
\left[ 
\ud \widetilde{r}^i (t) \partial_{\widetilde{r}^i} + \frac{\ud \widetilde{r}^i (t) \ud \widetilde{r}^j (t)}{2} \partial_{\widetilde{r}^i} \partial_{\widetilde{r}^j} 
\right] \delta^{(D)} ({\bf x} - \widetilde{\bf r}(t)) + o ((\ud t)^{3/2})\nonumber \\
&= 
\left[ 
- \ud \widetilde{r}^i (t) \partial_{x^i} + \nu (\ud t) \nabla^2 
\right] \delta^{(D)} ({\bf x} - \widetilde{\bf r}(t)) + o ((\ud t)^{3/2}) \, ,
\end{align}
where $\ud \widetilde{\bf r} (t)$ is given by Eq.\ (\ref{eqn:fsde}).
The derived equation is known as the Fokker-Planck equation. 
By using the forward SDE in the derivation, this becomes 
\begin{eqnarray}
\partial_t \rho ({\bf x},t) = 
- \nabla \cdot \{ {\bf u}_+ ({\bf x},t ) \rho ({\bf x},t) \}  + \nu \nabla^2  \rho ({\bf x},t)  \, .
\label{eqn:ffp}
\end{eqnarray}
Another Fokker-Planck equation from the backward SDE is given by
\begin{eqnarray}
\partial_t \rho ({\bf x},t) = 
- \nabla \cdot \{ {\bf u}_- ({\bf x},t ) \rho ({\bf x},t) \} - \nu \nabla^2  \rho ({\bf x},t)  \, .
\label{eqn:bfp}
\end{eqnarray}
The forward and backward SDE's describe different aspects of the same stochastic trajectory and hence the above two Fokker-Planck equations must be equivalent. 
Therefore, the following condition should be satisfied:
\begin{eqnarray}
 {\bf u}_+ ({\bf x},t) =  {\bf u}_- ({\bf x},t) + 2\nu \nabla  \ln \rho ({\bf x},t) 
+ \frac{1}{\rho ({\bf x},t)} \nabla \times {\bf A} ({\bf x}, t)
\, ,
\end{eqnarray}
where  ${\bf A} ({\bf x},t)$ is an arbitrary vector function. 
The physical interpretation of this vector function ${\bf A} ({\bf x},t)$ is still an open question, but for simplicity, it is often set to zero, leading to
\begin{eqnarray}
 {\bf u}_+ ({\bf x},t) =  {\bf u}_- ({\bf x},t) + 2\nu \nabla \ln \rho ({\bf x},t) \label{eqn:cc}
\, .
\end{eqnarray}
This is called the consistency condition.

Using the consistency condition, the two Fokker-Planck equations are reduced to 
the same equation of continuity associated with the conservation of probability,
\begin{equation}
\begin{split}
\partial_t \rho ({\bf x},t) &= - \nabla \cdot {\bf J}({\bf x},t) \, ,\\
{\bf J} ({\bf x},t) &= \rho({\bf x},t) {\bf v}({\bf x},t) \, ,
\end{split}
\label{eqn:eoc}
\end{equation}
where the velocity parallel to the probability current ${\bf J}$ is defined by 
\begin{equation}
  {\bf v} ({\bf x},t) = \frac{{\bf u}_+ ({\bf x},t) + {\bf u}_- ({\bf x},t)}{2} \, . \label{eqn:ave-vel}  
\end{equation}

The consistency condition is a key property to obtain the uncertainty relations in this formulation.
In fact, multiplying the condition by $x^i$, the expectation value is calculated as  
\begin{eqnarray}
\int \ud^D {\bf x} \, \rho ({\bf x},t) \left\{ x^i u^j_- ({\bf x},t) - x^i u^j_+ ({\bf x},t) \right\} 
= 2\nu \delta_{ij} \, .
\end{eqnarray}
As is seen later, we set $\nu = \hbar/(2\uM)$ to reproduce the Schr\"{o}dinger equation in the SVM quantization. 
Setting $\nu =\hbar/(2\uM)$, 
this relation is reminiscent of the expectation value of the canonical commutation relation, 
suggesting a deep connection between the stochastic formalism and conventional quantum mechanics.
See also Ref.\ \cite{I_Biane_2010}.

\section{Stochastic Variational Method for Particle} \label{sec:svm_particle}

We consider the stochastic variation for single-particle systems where
the classical form of the Lagrangian is given by Eq.\ (\ref{eqn:cla-lag}).

\subsection{Stochastic Action and its Variation}
\label{sec:sto_action}

To define the Lagrangian for the stochastic process, we replace ${\bf r}(t)$ with $\widetilde{\bf r}(t)$.
This replacement is straightforward for the potential term, but non-trivial for the kinetic term. 
Due to the existence of two distinct mean time derivatives, $\uD_+$ and $\uD_-$, 
the classical kinetic term can be generalized in multiple ways.
In this work, we assume that the stochastic representation of the kinetic term is given by the most general
quadratic form of the two derivatives,
\begin{eqnarray}
\frac{\uM}{2} \left[
A  (\uD_+ \widetilde{\bf r}(t) )^2 + B  (\uD_- \widetilde{\bf r}(t) )^2 
+ C (\uD_+ \widetilde{\bf r}(t) ) \cdot (\uD_- \widetilde{\bf r}(t) )
\right] \, ,
\end{eqnarray} 
where $A$, $B$ and $C$ are real constants. 
We require that SVM reproduces the result of the classical variation in the vanishing limit of the stochasticity. 
In this limit, both of $\uD_+ \widetilde{\bf r}(t)$ and $\uD_- \widetilde{\bf r}(t)$ coincide with the standard time derivative of a smooth trajectory. 
Thus, to reproduce the result of the classical variation in the limit, the three coefficients should satisfy 
\begin{eqnarray}
A + B + C = 1 \, .
\end{eqnarray}
The classical kinetic term, $\uM \dot{\bf r}^2(t)/2$, is therefore generalized to its stochastic counterpart \cite{U_Goncalves_2020},
\begin{eqnarray}
%\frac{\uM}{2} \dot{\bf r}^2(t)  
%\rightarrow 
\frac{\uM}{2} 
\left[
B_+ \sum_{l=\pm} A_l (\uD_l \widetilde{\bf r}(t) )^2 + B_- (\uD_+ \widetilde{\bf r}(t) ) \cdot (\uD_- \widetilde{\bf r}(t) )
\right] \, , \label{eqn:quad_rep}
\end{eqnarray}
where 
\begin{equation}
\begin{split}
A_\pm &= \frac{1}{2} \pm \alpha_A \, ,\\
B_\pm &= \frac{1}{2} \pm \alpha_B \, ,
\end{split}
\end{equation}
with $\alpha_A$ and $\alpha_B$ being real constants.
Equation (\ref{eqn:quad_rep}) in the vanishing limit of $\nu$ coincides with the classical kinetic term because
the two time derivatives coincide with the time derivative of a smooth trajectory.

In the classical variation, the form of a trajectory is entirely controlled for a given velocity. 
In SVM, however, we cannot determine a trajectory without ambiguity even for a given velocity field due to the stochasticity 
of the trajectory. 
Therefore we should consider the optimization of the averaged behavior of an action.
The action is eventually given by the expectation value, 
\begin{eqnarray}
I_{sto} [\widetilde{\bf r}] 
= \int^{t_f}_{t_i} \ud t \, \uE[L_{sto}(\widetilde{\bf r},\uD_+ \widetilde{\bf r}, \uD_- \widetilde{\bf r})]\, ,
\label{eqn:sto_act}
\end{eqnarray}
where the stochastic Lagrangian is defined by
\begin{eqnarray}
L_{sto} (\widetilde{\bf r},\uD_+ \widetilde{\bf r}, \uD_- \widetilde{\bf r}) 
= 
\frac{\uM}{2} (\uD_+ \widetilde{\bf r}(t), \uD_- \widetilde{\bf r}(t)) 
{\cal M} 
\left(
\begin{array}{c}
\uD_+ \widetilde{\bf r}(t) \\
\uD_- \widetilde{\bf r}(t)
\end{array}
\right) - V(\widetilde{\bf r}(t))
\, , \label{eqn:sto-lag}
\end{eqnarray}
with
\begin{eqnarray}
{\cal M} = 
\left(
\begin{array}{cc}
A_+ B_+ & \frac{1}{2} B_- \\
\frac{1}{2} B_-  & A_- B_+ 
\end{array}
\right) \, . \label{eqn:cal_m}
\end{eqnarray}

For the variation of the stochastic trajectory, instead of Eq.\ (\ref{eqn:cla-vari}), we consider 
\begin{eqnarray}
\widetilde{\bf r}(t) \longrightarrow \widetilde{\bf r}^\prime (t) = \widetilde{\bf r}(t) + \delta {\bf f} (\widetilde{\bf r}(t),t) \, .\label{eqn:csto-vari}
\end{eqnarray}
The infinitesimal smooth function $\delta {\bf f} ({\bf x},t)$ satisfies the same properties defined in Sec. \ref{sec:clas}, 
$\delta {\bf f}({\bf x},t_i) = \delta {\bf f}({\bf x},t_f) = 0$. 
Note that the fluctuation of $\delta {\bf f} (\widetilde{\bf r}(t),t) $ comes from that of $\widetilde{\bf r}(t)$.

The key feature of this stochastic variation appears when varying the kinetic term, 
which involves the mean derivatives. 
This requires a careful application of the stochastic integration by parts formula \eqref{eqn:spif}. For instance, the variation of the $(\uD_+ \widetilde{\bf r})^2$ term is calculated as follows:
\begin{align}
\int^{t_f}_{t_i} \ud t \,
\uE\left[
(\uD_+ \widetilde{\bf r}^\prime (t) )^2 
\right] 
&=
2  \int^{t_f}_{t_i} \ud t \,
\uE\left[ 
{\bf u}_+ (\widetilde{\bf r} (t),t ) \cdot \uD_+ \delta {\bf f} (\widetilde{\bf r}(t),t)
\right] \nonumber \\
&=
- 2  \int^{t_f}_{t_i} \ud t \,
\uE\left[ 
\uD_- {\bf u}_+ (\widetilde{\bf r} (t),t ) \cdot  \delta {\bf f} (\widetilde{\bf r}(t),t)
\right] \nonumber \\
&=
- 2  \int^{t_f}_{t_i} \ud t \,
\uE\left[ 
(\partial_t + {\bf u}_- (\widetilde{\bf r}(t),t) \cdot \nabla - \nu \nabla^2)  {\bf u}_+ (\widetilde{\bf r} (t),t ) \cdot  \delta {\bf f} (\widetilde{\bf r}(t),t)
\right] \, . 
\end{align}
From the second to the third line, we used the stochastic partial integration formula (\ref{eqn:spif}).
In the last line, It\^o's lemma is utilized.

On the other hand, the variation of the terms independent of time derivatives is the same as 
that in the corresponding classical variation.
Summarizing these results, the variation of the stochastic action is calculated as
\begin{eqnarray}
\delta I_{sto} [\widetilde{\bf r}]
=
 \int^{t_f}_{t_i} \ud t \, 
\uE\left[
{\bf I}_\delta (\widetilde{\bf r}(t),t)
 \cdot \delta {\bf f} (\widetilde{\bf r}(t),t)
\right] 
\label{eqn:deltai1}
\, ,
\end{eqnarray}
where
\begin{align}
{\bf I}_\delta ({\bf x},t) 
&= 
-\uM B_+ \left\{ A_+ (\partial_t + {\bf u}_- ({\bf x},t) \cdot \nabla - \nu \nabla^2)  {\bf u}_+ ({\bf x} ,t ) 
 + A_-  (\partial_t + {\bf u}_+ ({\bf x} ,t ) \cdot \nabla + \nu \nabla^2)  {\bf u}_- ({\bf x} ,t ) \right\}
\nonumber \\
& -
\uM \frac{B_- }{2} 
\left\{  (\partial_t + {\bf u}_- ({\bf x} ,t ) \cdot \nabla - \nu \nabla^2)  {\bf u}_- ({\bf x} ,t ) 
+ 
 (\partial_t + {\bf u}_+ ({\bf x} ,t ) \cdot \nabla + \nu \nabla^2)  {\bf u}_+ ({\bf x} ,t )  \right\} 
\nonumber \\
& - \nabla V ({\bf x}) \, .
\end{align}
In contrast to the classical case where the variation must vanish for any virtual displacement along a single trajectory, $\delta {\bf f}$, in SVM we require the expectation of the variation to vanish. 
By averaging over the ensemble of all possible initial positions, 
Eq.\ \eqref{eqn:deltai1} can be rewritten as:
\begin{eqnarray}
\int \ud^D {\bf R} \, \rho_0 ({\bf R})\delta I_{sto} [\widetilde{\bf r}]
=
  \int^{t_f}_{t_i} \ud t \int \ud^D {\bf x} \, \rho({\bf x},t)\, 
{\bf I}_\delta ({\bf x},t)
 \cdot \delta {\bf f} ({\bf x},t) =0
\label{eqn:deltai2}
\, ,
\end{eqnarray}
Since this must hold for any choice of initial probability distribution $\rho_0 ({\bf R})$ (and thus for any resulting $\rho({\bf x},t)$), 
and for any virtual displacement field $\delta {\bf f} ({\bf x},t)$, 
${\bf I}_\delta ({\bf x},t)$ must vanish identically for all ${\bf x}$ and $t$. This leads to the field equation:
${\bf I}_\delta ({\bf x},t) =0$, 
\begin{eqnarray}
\left(\partial_t + {\bf u}_- ({\bf x},t) \cdot \nabla - \nu \nabla^2, \,  \partial_t + {\bf u}_+ ({\bf x},t) \cdot \nabla + \nu \nabla^2 \right)
{\cal M} 
\left(
\begin{array}{c}
{\bf u}_+ ({\bf x},t ) \\
{\bf u}_- ({\bf x},t ) 
\end{array}
\right)
= - \frac{1}{\uM}\nabla V ({\bf x}) \, . \label{eqn:vari-particle1}
\end{eqnarray}
This is the stochastic equation of motion for the velocity fields. It correctly reduces to the classical Newton's equation in the vanishing noise limit ($\nu \rightarrow 0$). Therefore, SVM can be seen as a natural generalization of the classical variational method.

It is worth mentioning that the above result of the stochastic variation is formally represented by 
\begin{eqnarray}
\left[ \uD_+ \frac{\partial L_{sto}}{\partial (\uD_- \widetilde{\bf r})} + \uD_- \frac{\partial L_{sto}}{\partial (\uD_+ \widetilde{\bf r})} - \frac{\partial L_{sto}}{\partial \, \widetilde{\bf r}}
\right]_{\widetilde{\bf r}({\bf R},t) = {\bf x}} = 0 \, .
\label{eqn:sel}
\end{eqnarray}
This is the stochastic Euler-Lagrange equation. 
For the special case where $(\alpha_A,\alpha_B) = (0,1/2)$, 
the stochastic Euler-Lagrange equation coincides with the equation obtained by Nelson in Ref.\ \cite{D_Nelson_1966}.

\subsection{Schr\"{o}dinger Equation}

To demonstrate a key application of SVM, we now derive the Schr\"{o}dinger equation \cite{S_Yasue_1981}.
For this purpose, we choose the parameters $(\alpha_A, \alpha_B) = (0, 1/2)$, which gives the matrix ${\cal M} = {\rm diag} (1/2,1/2)$.
Then, the stochastic equation of motion (\ref{eqn:vari-particle1}) simplifies to
\begin{eqnarray}
 (\partial_t + {\bf v}({\bf x},t) \cdot \nabla) {\bf v}({\bf x},t) = - \frac{1}{\uM} \nabla ( V({\bf x}) + V_Q ({\bf x},t) ) \, ,
\label{eqn:vari_qh}
\end{eqnarray}
where the term $V_Q$ is identified as the quantum potential, defined by
\begin{eqnarray}
V_Q ({\bf x},t)= -  2 \uM \nu^2 \rho^{-1/2}({\bf x},t) 
\nabla^2 \sqrt{\rho ({\bf x},t)} \, .
\end{eqnarray}
To obtain this expression, we have used the averaged velocity (\ref{eqn:ave-vel}) and the consistency condition (\ref{eqn:cc}) to eliminate ${\bf u}_\pm ({\bf x},t)$.
The evolution of the probability density $\rho({\bf x},t)$ is given by the continuity equation (\ref{eqn:eoc}).
Thus, the SVM optimization for the single-particle Lagrangian leads to the coupled equations (\ref{eqn:eoc}) and (\ref{eqn:vari_qh}).

These coupled equations for $\rho$ and ${\bf v}$ can be unified into a single equation for a complex function.
To see this, we first assume the current velocity is irrotational and introduce an adimensional velocity potential $\theta ({\bf x},t)$, such that
\begin{eqnarray}
{\bf v}({\bf x},t) = 2\nu \nabla \theta ({\bf x},t) \, . \label{eqn:phase}
\end{eqnarray}
The stochastic equation of motion (\ref{eqn:vari-particle1}) can then be re-expressed as
\begin{eqnarray}
\partial_t \theta ({\bf x},t) + \nu (\nabla \theta ({\bf x},t) )^2 = - \frac{1}{2 \nu \uM} V ({\bf x}) + \nu\rho^{-1/2}({\bf x},t) \nabla^2 \sqrt{\rho ({\bf x},t)} \, .
\end{eqnarray}
We further introduce a complex function $\Psi ({\bf x},t)$ defined by
\begin{eqnarray}
\Psi ({\bf x},t) = \sqrt{\rho ({\bf x},t)} e^{\ii \theta ({\bf x},t)} \, , \label{eqn:wf}
\end{eqnarray}
which automatically satisfies the condition $|\Psi ({\bf x},t)|^2 = \rho ({\bf x},t)$.
The evolution equation for $\Psi ({\bf x},t)$ is found by combining the equations for $\rho({\bf x},t)$ and $\theta({\bf x},t)$, yielding
\begin{eqnarray}
\ii \partial_t \Psi ({\bf x},t) = \left[ -\nu \nabla^2 + \frac{1}{2\nu \uM} V({\bf x}) \right]\Psi ({\bf x},t) \, .
\end{eqnarray}
This becomes the Schr\"{o}dinger equation upon choosing the diffusion constant as
\begin{eqnarray}
\nu = \frac{\hbar}{2\uM} \, , \label{eqn:nu=hbar}
\end{eqnarray}
and $\Psi ({\bf x},t)$ is then identified with the wave function.
Indeed, Eq.\ (\ref{eqn:vari_qh}) is the hydrodynamical representation of the Schr\"{o}dinger equation, first proposed by Madelung \cite{Q_Madelung_1927} and extensively studied by Bohm and his collaborators \cite{A_Bohm_1952,Holland1995,A_Benseny_2014}.
From this quantum hydrodynamical perspective, all quantum effects are encapsulated in the quantum potential $V_Q ({\bf x},t)$.

From the SVM perspective, the quantization of classical systems is reinterpreted as a variational problem with higher microscopic resolution.
When observing phenomena on a macroscopic scale where the trajectory's non-differentiability is negligible, the classical variational principle applies, yielding Newton's equation.
Conversely, at microscopic scales where non-differentiability is significant, SVM must be employed, leading to the Schr\"{o}dinger equation.

A standard postulate in quantum mechanics is that the wave function must be a continuous and single-valued function.
This condition is not explicitly built into the hydrodynamical formulation.
In fact, the quantum potential becomes singular at the nodes of the wave function, necessitating an additional condition to connect solutions across these singularities.
The standard procedure is to impose a Bohr-Sommerfeld type quantization condition on a closed loop path of the quantum fluid. 
See Refs.\ \cite{Holland1995,U_Gazeau_2020,O_Takabayasi_1952,O_Wallstrom_1989} for details. 
It is worth mentioning that the quantum hydrodynamics formulation has an advantage in discussing quantum behaviors in generalized coordinates \cite{U_Gazeau_2020}.

\section{Symmetry and Stochastic Noether theorem}  \label{sec:s-noether}

A central pillar of modern physics is the profound connection between symmetries and conservation laws, formalized by Noether's theorem. A powerful feature of SVM is that this connection is preserved in a stochastic form \cite{N_MISAWA_88}. If the stochastic action is invariant under a continuous symmetry transformation, a corresponding physical quantity is conserved on average.

To illustrate this principle, we examine the consequences of spatial translation invariance for a free particle, where $V({\bf x}) = 0$. The stochastic Lagrangian is given by
\begin{equation}
L_{sto} (\widetilde{\bf r}, \uD_+ \widetilde{\bf r} , \uD_- \widetilde{\bf r})
= \frac{\uM}{4} \left( (\uD_+ \widetilde{\bf r}(t))^2 + (\uD_- \widetilde{\bf r}(t))^2 \right) \,,
\end{equation}
which corresponds to the choice $(\alpha_A,\alpha_B) = (0,1/2)$ in Eq.\ (\ref{eqn:sto-lag}).
This Lagrangian is manifestly invariant under an infinitesimal spatial translation,
\begin{equation}
\widetilde{\bf r}(t) \longrightarrow \widetilde{\bf r}^\prime (t) = \widetilde{\bf r} (t) + {\boldsymbol \varepsilon}\,, 
\end{equation}
where ${\boldsymbol \varepsilon}$ is an infinitesimal constant vector. The invariance of the action, $\delta I = 0$, under this transformation leads to a conservation law. The variation of the action can be expressed as a total time derivative:
\begin{align}
\delta I_{sto} &= \int^{t_f}_{t_i} \ud t \, \uE \left[ \frac{\partial L_{sto}}{\partial (\uD_+ \widetilde{\bf r})}\cdot \uD_+ {\boldsymbol \varepsilon} + \frac{\partial L_{sto}}{\partial (\uD_- \widetilde{\bf r})}\cdot \uD_- {\boldsymbol \varepsilon} \right] \nonumber \\
&= \int^{t_f}_{t_i} \ud t \, \frac{\ud}{\ud t} \uE \left[ \left( \frac{\partial L_{sto}}{\partial (\uD_+ \widetilde{\bf r})} + \frac{\partial L_{sto}}{\partial (\uD_- \widetilde{\bf r})} \right) \cdot {\boldsymbol \varepsilon} \right]\,.
\end{align}
The final equality is obtained by applying the stochastic Euler-Lagrange equation and the stochastic integration by parts formula.

Since $\delta I_{sto} = 0$ for an arbitrary ${\boldsymbol \varepsilon}$, the quantity inside the time derivative must be conserved. This gives the conserved Noether charge associated with spatial translation symmetry:
\begin{equation}
{\bf P}(t) 
= \uE \left[ \frac{\partial L_{sto}}{\partial (\uD_+ \widetilde{\bf r})} + \frac{\partial L_{sto}}{\partial (\uD_- \widetilde{\bf r})} \right] 
= \frac{\uM}{2} \uE \left[ {\bf u}_+ (\widetilde{\bf r}(t),t) + {\bf u}_- (\widetilde{\bf r}(t),t) \right]  \, .
\end{equation}
Using the definition of the probabilistic velocity ${\bf v}$, this conservation law is expressed as
\begin{equation}
\frac{\ud}{\ud t} \uE[\uM {\bf v}(\widetilde{\bf r}(t),t)] = 0\,.
\end{equation}

This conserved Noether charge is not just an abstract quantity; it is precisely the quantum mechanical expectation value of the momentum operator. By using the definitions of the current velocity ${\bf v}$ and the wave function $\Psi$, this charge is shown to be the quantum-mechanical expectation value of momentum; 
\begin{align}
\uE[\uM {\bf v}] = \int \ud^D {\bf x} \, \uM {\bf v} ({\bf x},t) \rho ({\bf x},t) 
= \int \ud^D {\bf x} \, \Psi^* ({\bf x},t)  (- \ii \hbar \nabla) \Psi ({\bf x},t) = \langle - \ii \hbar \nabla \rangle \,.
\label{eqn:momentum-exp}
\end{align}
This correspondence is not limited to spatial translations. Invariances under time translation and rotations similarly lead to the conservation of the expectation values of energy and angular momentum, respectively \cite{N_MISAWA_88}. Thus, the stochastic Noether theorem provides a physically intuitive and rigorous foundation for the conservation laws observed in quantum mechanics, rooting them in the fundamental symmetries of the system's Lagrangian.

\section{Numerical Validation: The Double-Slit Experiment} \label{sec:numerical}

To provide a concrete validation of the formalism, we demonstrate that a numerical simulation of the underlying stochastic process correctly reproduces a typical quantum phenomenon: the double-slit interference pattern. The consistency between the two approaches is established by showing that the probability distribution generated by an ensemble of Brownian trajectories matches the probability density $|\Psi|^2$ obtained from the Schr\"{o}dinger equation.

We consider a standard double-slit setup, simplified by assuming a homogeneous flux in the forward ($x$) direction and analyzing the dynamics in the transverse ($y$) dimension. The initial wave function at $t=0$ is a superposition of two Gaussian distributions centered at the two slits, separated by a distance $2d$:
\begin{equation}
\Psi(y,0) \propto \exp\left[-2 \left(\frac{y}{d}-1\right)^2 \right] + \exp\left[-2 \left(\frac{y}{d}+1\right)^2 \right] \, . \label{eqn:ini_wf}
\end{equation}
The exact time evolution of this wave function can be obtained by solving the free Schr\"{o}dinger equation, which yields
\begin{align}
\Psi (y,t)
&=
\frac{2}{(2\pi \sigma^2_t)^{1/4}} 
\exp\left\{i \left[
-\frac{\alpha_t}{2} 
+  \frac{\hbar t (y^2 + d^2)}{\uM d^2 \sigma^2_{t}}
\right]\right\} 
\exp\left\{-\frac{y^2 + d^2}{4 \sigma^2_{t}}\right\}
\nonumber \\
& \times \sqrt{ \cosh^2 \left( C(y,t) \right)  \cos^2 \left( B(y,t)  \right) + \sinh^2 \left(  C(y,t) \right)  \sin^2  \left(  B(y,t)  \right) }\exp\left\{-i\gamma(y,t)\right\} \, ,
\end{align}
where the time-dependent parameters are defined as
\begin{align}
\sigma_t 
&=
\sigma_0 \sqrt{1 + \frac{16 \hbar^2 t^2}{\uM^2 d^4}} 
\, , \\
\alpha_t &= {\rm tan}^{-1}  \frac{4 \hbar t}{\uM d^2} \, ,\\
B(y,t)
&= 
y \frac{2 \hbar t d }{\uM d^2 \sigma^2_{t}} \, , \\
C(y,t)
&=
 y \frac{d}{2 \sigma^2_{t}}, \\
\gamma(y,t) 
&= \tan^{-1}( \tanh C(y,t) \tan B(y,t) ).
\end{align}
This analytical solution provides the exact quantum mechanical probability density $|\Psi(y,t)|^2$ that our simulation aims to reproduce.

The numerical simulation consists of generating an ensemble of stochastic trajectories governed by the forward SDE (\ref{eqn:fsde}). The crucial input for this simulation is the vector field ${\bf u}_+ ({\bf x},t)$, which guides the Brownian motion. Using the relations derived in the previous sections, this field is determined from the analytical wave function $\Psi$ as
\begin{align}
{\bf u}_+ ({\bf x},t) 
&= \frac{\hbar}{\uM} \nabla \theta({\bf x},t) + \frac{\hbar}{2\uM} \nabla \ln \rho ({\bf x},t) 
\nonumber \\
&= \frac{\hbar}{\uM} {\rm Im} \{ \nabla \ln \Psi({\bf x},t) \} + \frac{\hbar}{\uM} \nabla \ln |\Psi({\bf x},t)|^2 
 \, .
\end{align}
The simulations are performed in a special unit system where $\hbar = 1$ and $\uM = 1$. The time is normalized by $\tau_0 = \uM d^2 / \hbar$.

\begin{figure}[h!]
\centering
\includegraphics[scale=0.55]{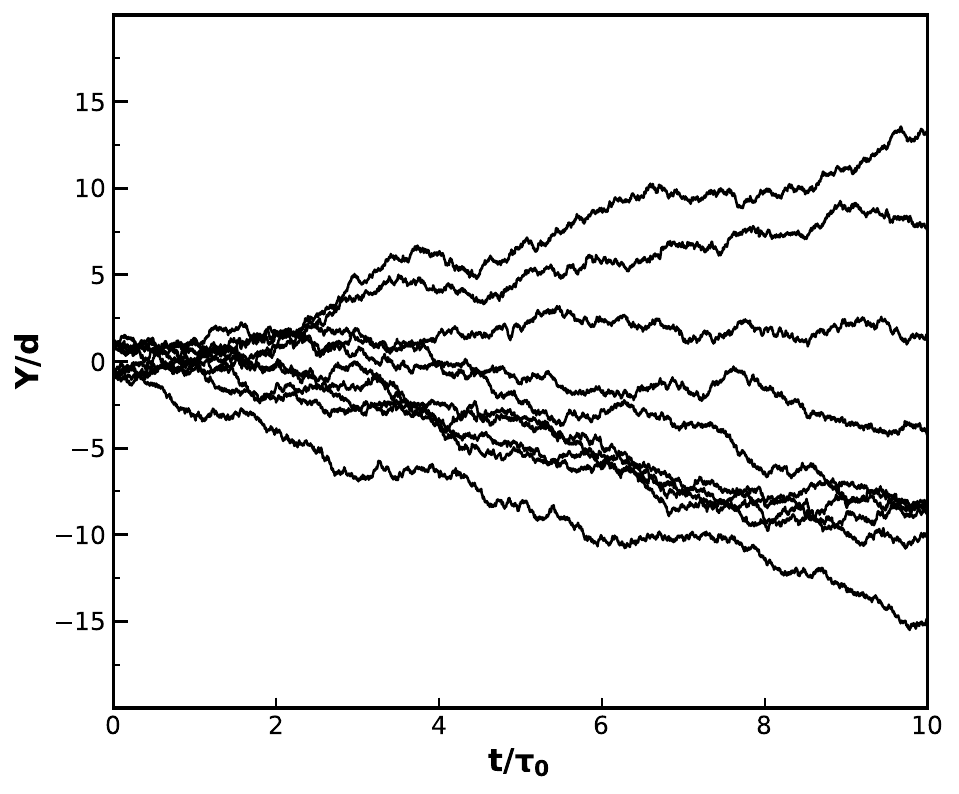}
\caption{Sample stochastic trajectories generated by the numerical simulation of the double-slit experiment within the framework of SVM. The trajectories represent the zigzag motion of Brownian particles governed by the forward stochastic differential equation (\ref{eqn:fsde}). The velocity field for the simulation is derived from the analytical solution to the free Schr\"{o}dinger equation with an initial wave function given by Eq.\ (\ref{eqn:ini_wf}), which describes two Gaussian wave packets centered at the slits ($y=\pm d$). Five sample trajectories are shown originating from the vicinity of each slit. The simulation uses units where $\hbar = 1$, $\uM = 1$, and time is measured with scale $\tau_0 = \uM d^2 / \hbar$. This figure is adapted from Fig.\ 1 of Ref.\ \cite{U_Goncalves_2024}.}
\label{fig:trajetorias}
\end{figure}

\begin{figure}[h!]
\includegraphics[scale=0.35]{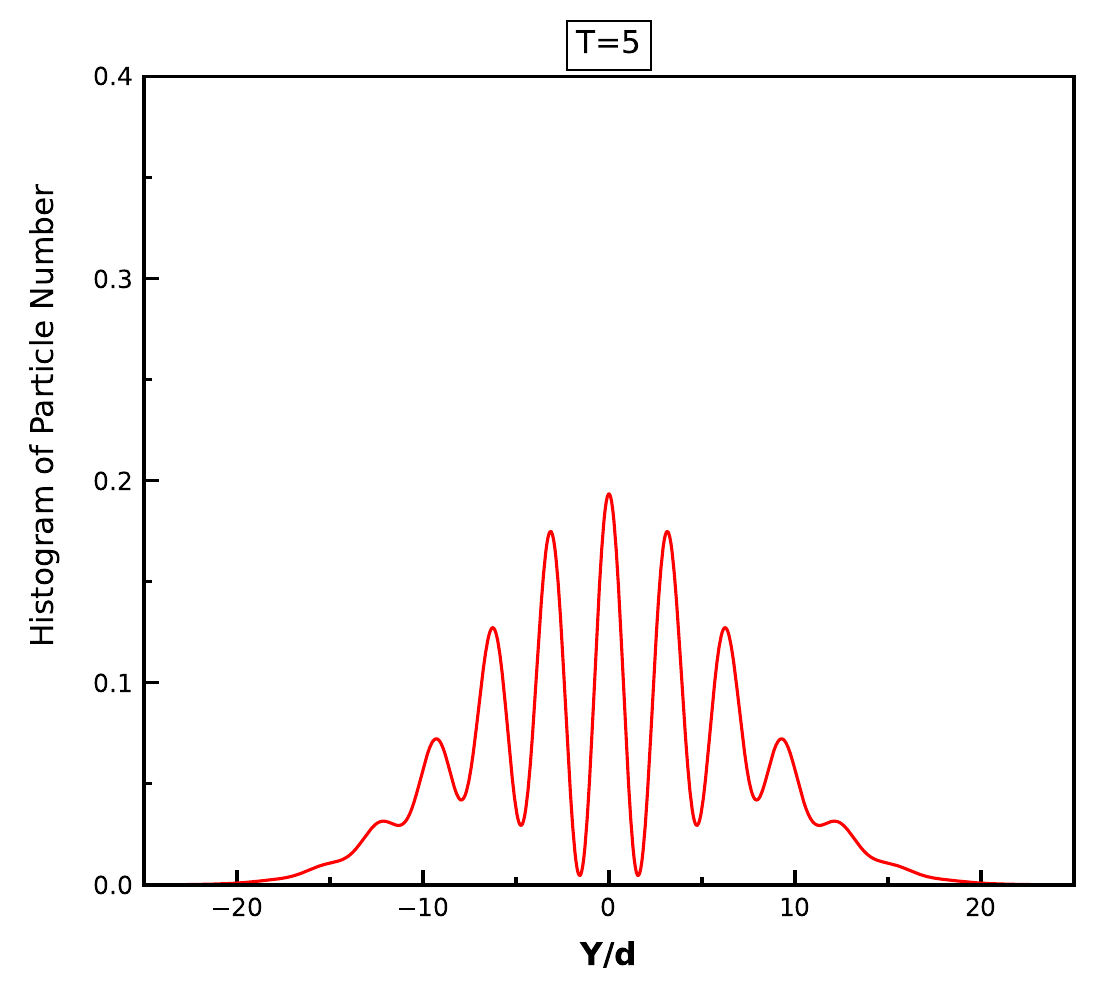}
\includegraphics[scale=0.35]{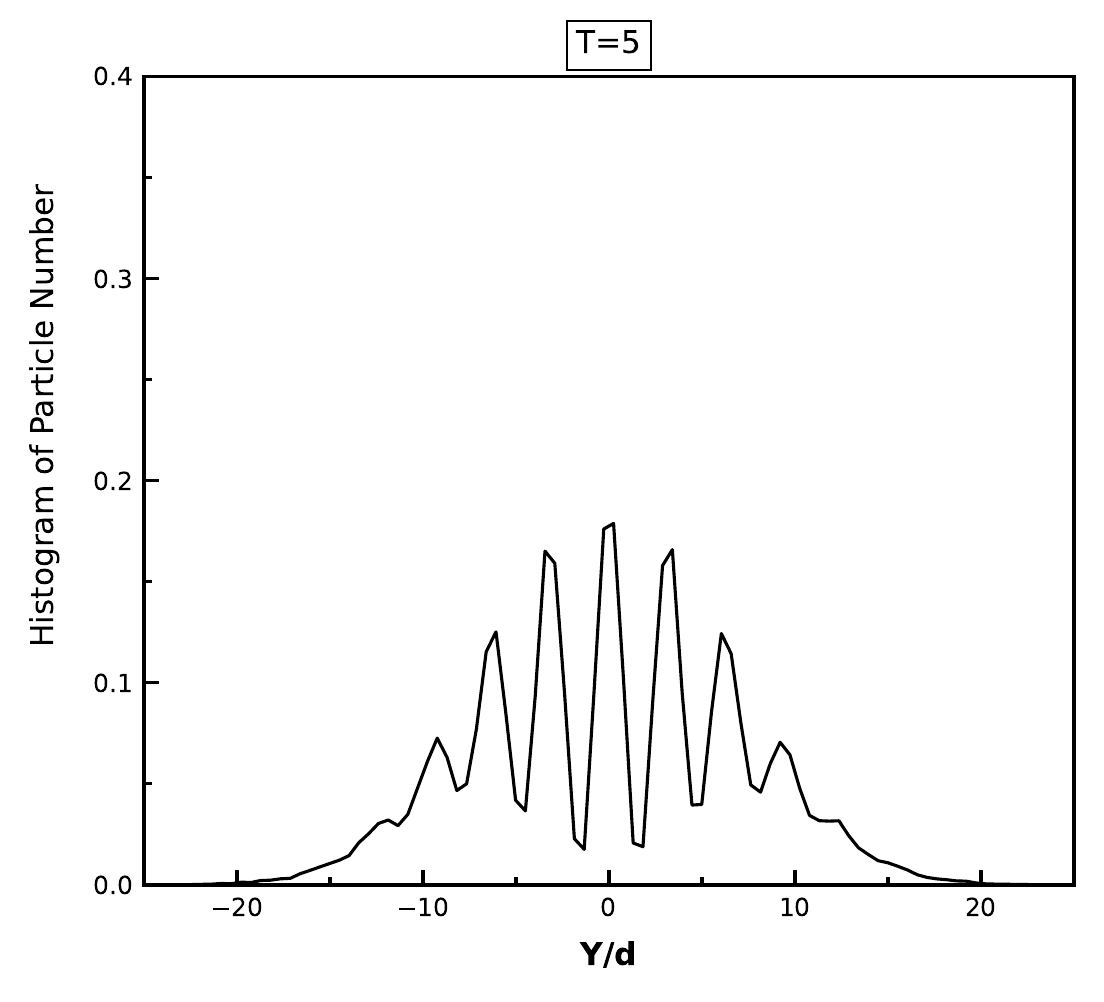}
\caption{Numerical validation of SVM by comparing the particle distribution in a double-slit experiment. 
\textbf{Left panel:} The exact quantum mechanical probability density $|\Psi(y,t)|^2$ at $T = t/ \tau_0 = 5$ with $\tau_0 = \uM d^2 / \hbar$, calculated from the analytical solution of the free Schr\"{o}dinger equation. The characteristic interference fringes are clearly visible. 
\textbf{Right panel:} The particle distribution at the same time $T = t/ \tau_0 = 5$ shown as a histogram, generated by numerically simulating 10,000 individual Brownian trajectories with $\ud t=0.001$. Each trajectory evolves according to the forward SDE (\ref{eqn:fsde}), guided by ${\bf u}_+({\bf x},t)$ derived from the wave function. The agreement between the histogram and the analytical curve demonstrates that the ensemble of stochastic trajectories correctly reproduces the quantum interference pattern, thus validating the SVM approach. These figures are adapted from Fig.\ 2 of Ref.\ \cite{U_Goncalves_2024}.}
\label{fig:simulacao}
\end{figure}

Figure \ref{fig:trajetorias} shows sample stochastic trajectories generated by the numerical simulation of the double-slit experiment within the framework of SVM. 
The trajectories represent the zigzag motion of Brownian particles governed by the forward SDE (\ref{eqn:fsde}). Five sample trajectories are shown originating from the vicinity of each slit. 
%The simulation uses units where $\hbar = 1$, $\uM = 1$, and time is measured with scale $\tau_0 = \uM d^2 / \hbar$.

The main result is presented in Fig.\ \ref{fig:simulacao}, which compares the particle distribution from the simulation with the exact quantum mechanical probability density at a given time $T = t/ \tau_0 = 5$ with $\tau_0 = \uM d^2 / \hbar$. 
The histogram in the right panel, generated from 10,000 simulation events with $\ud t/\tau_0 =0.001 $, shows the agreement with the analytical interference pattern shown in the left panel. 
This result numerically validates that the stochastic process defined by SVM correctly reproduces the predictions of quantum mechanics.

\section{Stochastic Processes in Curved Geometry with Torsion} \label{sec:svm_torsion}

In the SVM framework, quantum mechanics is formulated as a stochastic optimization of the classical action. 
Assuming this principle holds even in a general curved space, 
quantum mechanics in curved spaces can be formulated by extending the SVM formalism. SVM in curved space was first formulated in Refs.\ \cite{N_Koide_2019,V_Koide_2020}. 
A further extension to curved spaces with torsion was formulated 
in Ref.\ \cite{Armin_Koide_2025}.
The following discussion is based on the results of these works.
Before proceeding with our derivation, it is important to emphasize again our working assumption regarding the dynamical status of the geometry. In the following formulation, the spatial geometry, encompassing both the metric and the torsion, is treated strictly as a fixed classical background field. We do not solve the corresponding gravitational field equations for these geometric quantities, nor do we consider the backreaction of the quantum fluctuations on the background geometry.

\subsection{Forward SDE in Curved Geometry with Torsion}

Extending the SDEs from flat to curved space is a non-trivial task. 
The main difficulty is that the Wiener process implies non-tensorial transformation rules, and thus a standard definition of noise often violates the intrinsic geometry of the manifold.
%Generalizing the SDE from a flat to a curved space is a non-trivial task. 
%The primary difficulty arises because the Wiener process does not transform as a simple vector under general coordinate transformations, and a naive approach can violate the geometric constraints of the manifold.
%The core of this problem is that a naively defined noise term does not respect the geometric constraints of the coordinate system. 
In polar coordinates, for instance, the radial component $r(\widetilde{r})$ must always be positive. However, if the noise term in the radial direction were a simple Wiener process, its increment could take a large negative value, potentially forcing $r(t)$ to become unphysically negative during its time evolution.

To systematically address this issue within the SVM framework, we adopt the vielbein formalism.
This approach allows us to define the Wiener process in a local orthogonal frame, and then project it onto the curved geometry using the vielbein.
Using the standard Wiener process $\widetilde{B}^{\hat{a}}_+ (t)$, 
the forward SDE written in generalized coordinates $q^{i}$ is defined by 
\begin{equation}
d \widetilde{q}^i (t) = 
u^i_+ (\widetilde{q}(t),t) dt +\sqrt{2\nu} 
\tensor{e}{_{\hat{a}}^i} \circ \ud\widetilde{B}^{\hat{a}}_+ (t) \,\,\, (\ud t >0)
 \, . \label{eqn:curve_fSDE}
\end{equation}
Here $i (=1,\cdots,D)$ denotes the spatial components of generalized coordinates and the dual vielbein is evaluated at $\widetilde{q}(t)$; 
$\tensor{e}{_{\hat{a}}^i}:=\tensor{e}{_{\hat{a}}^i}(\widetilde{q}(t))$.
The standard Wiener process $\widetilde{B}^{\hat{a}}_+ (t)$ satisfies the same correlation properties given by Eq.\ (\ref{eqn:f_wiener}).
The symbol $``\circ"$ denotes the Stratonovich product defined for arbitrary smooth functions $F(q,t)$ and $G(q,t)$ as follows:
\begin{equation}
F(\widetilde{q}(t),t) \circ \ud G (\widetilde{q}(t),t) := 
\frac{F(\widetilde{q}(t+\ud t),t+\ud t)+F(\widetilde{q}(t),t)}{2} \ud G (\widetilde{q}(t),t) \, .
\end{equation}
In stochastic calculus, the Leibniz rule does not hold in general. 
However, by employing the Stratonovich definition, it can be formally restored \cite{Gardiner_2004}.

To evaluate the Stratonovich product in Eq.\ (\ref{eqn:curve_fSDE}), 
we must define how the vielbein 
$\tensor{e}{_{\hat{a}}^i}$ changes along the stochastic trajectory $\widetilde{q}(t)$.
In standard differential geometry \cite{Kobayashi2009}, the change of a frame along a smooth trajectory is defined by the horizontal lift, which provides the intuitive picture of the frame "rolling without slipping" on the manifold. The generalization of this concept to stochastic trajectories is known as stochastic parallel transport, first introduced by It\^{o} \cite{S_Ito_1975}. Following this approach, the stochastic change of the vielbein is defined as:
\begin{equation} 
\tensor{\ud e}{_{\hat{a}}^i} 
:= -  {\Gamma^{i}}_{j k}  \tensor{e}{_{\hat{a}}^j}  \circ  \ud \widetilde{q}^k(t) \, . \label{eqn:de}
\end{equation}
See also, Sec. V of Ref. \cite{Ikeda2014} and Sec. 2.3 of Ref.\ \cite{Hsu_2002}. 
Using this, Eq.\ (\ref{eqn:curve_fSDE}) is reexpressed as
\begin{equation}    
\ud \widetilde{q}(t) 
= 
U^{i}_+ (\widetilde{q}(t),t) dt + \sqrt{2\nu}  \tensor{e}{_{\hat{a}}^i} \ud \widetilde{B}^{\hat{a}}_+ (t) \, ,
\end{equation}
where
\begin{equation}
U^{i}_+ (\widetilde{q}(t),t) 
:= u^i_+ (\widetilde{q}(t),t) - \nu 
{\Gamma^{i}}_{jk} \, g^{jk} 
 \, .
\end{equation}
The final term in the definition of $U^{i}_+$, namely $-\nu {\Gamma^{i}}_{jk} g^{jk}$, is the so-called It\^{o}-Stratonovich correction term. This term is not an arbitrary addition but emerges naturally from the covariant formulation to ensure the stochastic process is consistent with the underlying geometry.
The presence of this term ensures that the stochastic process behaves properly on the manifold, 
for example, by keeping the radial coordinate $r$ positive in polar coordinates.

The particle probability distribution in curved geometry is defined by 
\begin{equation}
\rho_{den} ({q},t)= \int \sqrt{g(Q)}d^D Q \, \rho_0 (Q) E[\delta^{(D)} (q-\widehat{q}(t))] \, ,
\end{equation}
with $g = {\rm det} (g_{ij})$, $Q = \widetilde{q}(t_i)$ and $\rho_0 (Q)$ being the initial probability distribution.
The corresponding scalar distribution is defined by
\begin{equation}
\rho ({q},t)= \frac{1}{\sqrt{g({q})} }\int \sqrt{g(Q)} d^D Q \, \rho_0 (Q) E[\delta^{(D)} (q-\widetilde{q}(t))] \, .
\end{equation}
This satisfies the following Fokker-Planck equation:
\begin{equation}
\partial_t \rho (q,t) 
= 
-\nabla_j  \left(\rho(q,t)
\left\{
u^j_+ (q(t),t)  + \nu K^{j}
\right\} \right) 
+ \nu \Delta  \rho (q,t) 
\, . \label{eqn:fpeq_torsion}
\end{equation}
where $\nabla_i$ is the covariant derivative, $\Delta := g^{ij} \nabla_i \nabla_j$ 
is the connection Laplacian. 
To define the contracted torsion tensor $K^{i}$, we introduce the contorsion tensor $K\indices{^{k}_{ij}}$ which is defined in terms of the torsion tensor $S\indices{^{k}_{ij}}$: 
\begin{equation}
K^{i} := g^{ij} K_j = g^{ij} {K^k}_{j k} = 2 g^{ij} {S^k}_{jk}  \, ,
\end{equation}
with 
\begin{align}
K\indices{^{k}_{ij}} 
&\coloneqq 
    g\indices{^{kl}} \left( S\indices{_{lij}} - S\indices{_{ilj}} - S\indices{_{jli}} \right) \, ,\\
S\indices{^{k}_{ij}} 
&\coloneqq \Gamma\indices{^{k}_{[ij]}} = \frac{1}{2} \left( \Gamma\indices{^{k}_{ij}} - \Gamma\indices{^{k}_{ji}} \right) \, .
\end{align}
We will explore how this additional degree of freedom, represented by torsion, influences quantum dynamics.
In this derivation we used, 
\begin{equation}
\frac{1}{\sqrt{g}} \partial_i \sqrt{g} 
= {\Gamma^k}_{k i} = \mathring{\Gamma^k}_{k i}  \, ,
\end{equation}
and
\begin{align}
 \partial_k g^{ik} \sqrt{g}
&= 2 g^{il} \sqrt{g} {S^k}_{kl} - 
g^{lm} \sqrt{g}
{\Gamma^{i}}_{lm} \, .
\end{align}

\subsubsection{Example: Brownian Motion in Polar Coordinates}

As an example, let us apply the above definition to describe Brownian motion in flat polar coordinates $(x=r\cos \theta, y=r\sin \theta)$, 
where the vielbein and its dual are defined by 
\begin{align}
\begin{array}{cc}
 \tensor{e}{_{\hat{x}}^r} 
 = \cos \theta \, ,& 
 \tensor{e}{_{\hat{y}}^r} 
 = \sin \theta \, , \\
 \tensor{e}{_{\hat{x}}^\theta} 
 = - \frac{\sin \theta}{r} \, ,&
 \tensor{e}{_{\hat{y}}^\theta} 
 = \frac{\cos \theta}{r} \, ,
 \end{array} 
 \\
 \begin{array}{cc}
 \tensor{e}{^{\hat{x}}_r}
 = \cos \theta \, , &
 \tensor{e}{^{\hat{x}}_\theta}
 = - r\sin \theta \, ,\\
  \tensor{e}{^{\hat{y}}_r}
= \sin \theta \, , &
 \tensor{e}{^{\hat{y}}_\theta}
= r \cos \theta \, .
\end{array}
\end{align}
Using Eq.\ (\ref{eqn:curve_fSDE}), the forward SDE's for polar coordinates are given by 
\begin{align}
\ud \widetilde{r}(t)
&
=
\sqrt{2\nu} 
\tensor{e}{_{\hat{a}}^r} \circ \ud\widetilde{B}^{\hat{a}}_+ (t)
= \frac{\nu}{\tilde{r}(t)} \ud t + \sqrt{2\nu} \ud \widetilde{B}^r (t) \label{eqn:SDE_polar_r}\, ,\\
\ud \widetilde{\theta}(t)
&
=
\sqrt{2\nu} 
\tensor{e}{_{\hat{a}}^\theta} \circ \ud\widetilde {B}^{\hat{a}}_+ (t)
= \frac{\sqrt{2\nu}}{\widetilde{r}(t)} \ud \widetilde{B}^\theta (t) \, ,
\end{align}
where we introduced new noise terms which are expressed in terms of the standard Wiener process $\widetilde{B}^x (t)$ and $\widetilde{B}^y (t)$,  
\begin{align}
\ud \widetilde{B}^r(t) 
&:= \cos \widetilde{\theta} (t) \ud \widetilde{B}^x (t) + \sin \widetilde{\theta} (t) \ud \widetilde{B}^y (t) \, , \label{eqn:BMpolar_r} \\
\ud \widetilde{B}^\theta (t) 
&:= -\sin \widetilde{\theta} (t) \ud \widetilde{B}^x (t) + \cos \widetilde{\theta} (t) \ud \widetilde{B}^y (t) \, ,
\end{align}
where $\ud \widetilde{B}^r(t)$ and $\ud \widetilde{B}^\theta(t)$ are the projections of the Cartesian Wiener process onto the local polar basis vectors.
The first term on the right-hand side of Eq.\ (\ref{eqn:SDE_polar_r}), prevents $\tilde{r}(t)$ from taking negative values.

One can verify that the Fokker-Planck equation corresponding to the forward SDEs is indeed the standard diffusion equation in polar coordinates:
\begin{equation}
    (\partial_t  - \nu \nabla^2 )\rho(r,\theta,t) = 0 \, .  
\end{equation}
This confirms that our covariant SDE formalism correctly describes Brownian motion.

\subsection{Backward SDE and Consistency Condition in Curved Geometry with Torsion}

To formulate SVM, we need the backward SDE which describes the stochastic process equivalent to the forward SDE (\ref{eqn:curve_fSDE}), which is defined by 
\begin{equation}
d \widetilde{q}^i (t) = 
u^i_- (\widetilde{q}(t),t) dt + \sqrt{2\nu} e^i_{\hat{a}} (\widetilde{q}(t)) \circ d\widetilde{B}^{\hat{a}}_- (t) \,\,\,  (\ud t < 0) \, , \label{eqn:bSDE}
\end{equation}
where the correlation properties of $d\widetilde{B}^{\hat{a}}_- (t)$ are defined by Eq. (\ref{eqn:b_wiener}).
The above can be reexpressed without the Stratonovich definition, 
\begin{equation}
d \widetilde{q}^i (t) 
= 
U^{i}_- (\widetilde{q}(t),t) dt + \sqrt{2\nu}  e^i_{\hat{a}} (\widetilde{q}(t))  d\widetilde{B}^{\hat{a}}_- (t) \, ,
\end{equation}
where
\begin{equation}
U^{i}_- ({q},t) 
= u^i_- (\widehat{q}(t),t) + \nu 
{\Gamma^{i}}_{\alpha j}(\widehat{q}(t)) \, g^{\alpha j} (\widehat{q}(t)) 
 \, .
\end{equation}
In the backward SDE, the same It\^{o}-Stratonovich correction appears but the sign is 
opposite.

The corresponding Fokker-Planck equation for the scalar distribution is 
\begin{align}
\partial_t  \rho (q,t) 
&= 
-\nabla_j  \left(\rho(q,t)
\left\{
u^j_- (q(t),t) - \nu K^{j}
\right\} \right) 
- \nu  \Delta \rho (q,t) 
\, . \label{eqn:fpeq_torsion2}
\end{align}
This should be equivalent to the Fokker-Planck equation of the forward SDE (\ref{eqn:fpeq_torsion}), leading to the consistency condition for curved geometry with torsion, 
\begin{equation}
u^j_+ (q,t) - u^j_-(q,t) = -2 \nu K^j + 2 \nu g^{jk} \partial_k \ln \rho (q,t) \, .
\end{equation}
Using this, $u^i_{\pm}$ is expressed as 
\begin{align}
u^i_+  &= v^i  - \nu K^i + \nu g^{ij}\partial_j \ln \rho  \label{eqn:u+torsion}\, , \\
u^i_-  &= v^i  + \nu K^i - \nu g^{ij}\partial_j \ln \rho  \label{eqn:u-torsion}\, . 
\end{align}
where a new velocity field is defined by 
\begin{equation}
v^l (q,t):= \frac{u^l_+ (q,t)+ u^l_- (q,t)}{2} \, .\label{eqn:del_v_cur_tor}
\end{equation}
As is the case in flat geometry without torsion, the FP equations obtained from the forward and backward SDE's is reduced to the same equation of continuity,
\begin{align}
\partial_t \rho (q,t)
=- \nabla_j (\rho (q,t) v^{j} (q,t)) \, . \label{eqn:eofconti}
\end{align}

\section{Stochastic Variational Method with Torsion} \label{sec:svm_torsion2}

As a crucial preliminary step, let us first examine the variational principle in the presence of a vielbein field for a classical particle. We consider a non-relativistic particle of mass $\uM$ in a curved geometry with torsion. The classical Lagrangian of this system is assumed to be given by  
\begin{equation}
L = \frac{\uM}{2} g_{ij} \dot{q}^{i} \dot{q}^j - V \left( q \right) \, ,
\label{eqn:curve_cla_lag1}
\end{equation}
where $V(q)$ is a scalar function representing a potential energy and 
$\dot{q}^{i} = \ud q^{i}/ \ud t$.
The variation of this Lagrangian with respect to the generalized coordinates $q^i$ gives the well-known equation of motion:
\begin{equation}
\ddot{q}^i + \mathring{\Gamma^{i}}_{j k} \dot{q}^j \dot{q}^k 
= - \frac{1}{\uM} g^{ik} \partial_k V (q) \label{eqn:cla_eom_curve1}\, ,
\end{equation}
where $\ddot{q}^i = \ud^2 q^{i}/ \ud t^2$ and $\mathring{\Gamma^{i}}_{j k}$ is the Levi-Civita connection. For a free particle ($V=0$), this is the geodesic equation.

Alternatively, the same Lagrangian can be expressed in local orthogonal coordinates,
\begin{equation}
L = \frac{\uM}{2} \eta_{\hat{a}\hat{b}} \dot{q}^{\hat{a}} \dot{q}^{\hat{b}} - V \left( q \right) \, , \label{eqn:curve_cla_lag2}
\end{equation}
where $\dot{q}^{\hat{a}}$ is the velocity in the local frame and $\eta_{\hat{a}\hat{b}} ={\rm diag}(1,1,\cdots,1)$ satisfying $g_{ij} = \eta_{\hat{a}\hat{b}} \tensor{e}{^{\hat{a}}_i}\tensor{e}{^{\hat{b}}_j}$. 
The variation with respect to $q^{\hat{a}}$ gives the simpler equation:
\begin{equation}
\ddot{q}^{\hat{a}}= - \frac{1}{\uM}\eta^{\hat{a}\hat{b}} \partial_{\hat{b}} V  \, .
\end{equation}
A fundamental principle is that the resulting equation of motion must be independent of the coordinate system used to express the Lagrangian. To check this consistency, we transform the local equation of motion back to general coordinates. Using the relation $\dot{q}^i = \tensor{e}{_{\hat{a}}^i} \dot{q}^{\hat{a}}$ and assuming the horizontal lift of the vielbein is governed by the full affine connection $\Gamma$, we obtain:
\begin{align}
\ddot{q}^i 
&= \left(\frac{\ud}{\ud t} \tensor{e}{_{\hat{a}}^i}\right) \dot{q}^{\hat{a}} + \tensor{e}{_{\hat{a}}^i} \ddot{q}^{\hat{a}} \nonumber \\
&= - {\Gamma^{i}}_{j k}\dot{q}^k (\tensor{e}{_{\hat{a}}^j} \dot{q}^{\hat{a}}) - \frac{1}{\uM} g^{ik} \partial_k V \nonumber \\
&= - {\Gamma^{i}}_{j k}\dot{q}^k \dot{q}^{j} - \frac{1}{\uM} g^{ik} \partial_k V 
\label{eqn:cla_eom_curve2}\, .
\end{align}
At first glance, this result (\ref{eqn:cla_eom_curve2}) appears to contradict Eq.\ (\ref{eqn:cla_eom_curve1}), as it involves the full affine connection $\Gamma$ instead of the Levi-Civita connection $\mathring{\Gamma}$. 
This is a well-known fundamental issue, often referred to as the problem of disagreement between autoparallels and geodesics in theories with torsion \cite{Hehl1}.
While this inconsistency might be resolved by considering a non-minimal coupling of torsion in the Lagrangian, there is no universally established procedure for this, especially for spinless particles. 
Therefore, instead of modifying the Lagrangian, we impose a restriction on the property of torsion itself. 
The consistency between Eqs.\ (\ref{eqn:cla_eom_curve2}) and (\ref{eqn:cla_eom_curve1}) 
is achieved if the contortion tensor term $K\indices{^{i}_{jk}}\dot{q}^j \dot{q}^k$ vanishes. 
Since $\dot{q}^j \dot{q}^k$ is symmetric in $j,k$, 
this is satisfied if the torsion tensor $S\indices{^{i}_{jk}}$ is {\it totally antisymmetric}.
In this situation, the torsional contribution to the equation of motion disappears, 
and the two equations become identical. Consequently, in the remainder of this paper, 
we will exclusively focus on curved spaces with totally antisymmetric torsion.

This assumption of the totally antisymmetry is also plausible from 
a cosmological perspective. 
It is known that employing the cosmological principle together with
energy-momentum conservation 
and standard cosmological expansion restricts
the form of torsion to the case of a totally antisymmetric tensor \cite{vandeVenn2022,C_Tsamp_1979}.
The totally antisymmetric torsion is furthermore 
preferred from a MAG standpoint as it allows for a
continuous limit of vanishing torsion, thereby recovering GR \cite{C_Fabbri_2014}.
%This replacement is generally not justified. The difference between the two connections is the contorsion tensor, $K^i{}_{jk} \coloneqq \Gamma^i{}_{jk} - \mathring{\Gamma}^i{}_{jk}$. 
%The term in the equation of motion involves the contraction $K^i{}_{jk}\dot{q}^j \dot{q}^k$. Since $\dot{q}^j \dot{q}^k$ is symmetric in its indices, only the symmetric part of the contorsion contributes. 
%In a general geometry with torsion, 
%his symmetric part is non-zero.
%
%However, consistency between the two approaches can be achieved in the special case where the torsion tensor $S^i{}_{jk}$ is {\it totally antisymmetric}, the contorsion tensor becomes purely antisymmetric in its lower indices ($K^i{}_{jk} = -K^i{}_{kj}$), 
%and thus its symmetric part vanishes, $K^i{}_{(jk)}=0$. 
%In this situation, the torsional contribution to the equation of motion disappears, and Eq.\ (\ref{eqn:cla_eom_curve2}) reduces to Eq.\ (\ref{eqn:cla_eom_curve1}).
%In this review, we exclusively focus on curved geometry with totally antisymmetric torsion.

Note that the above equation can be regarded as the equation to describe the velocity field, 
\begin{equation}
\partial_t v^{i} (q,t) + v^{j} (q,t) \nabla_j v^{i} (q,t) 
= - \frac{1}{\uM} g^{ik} \partial_k V (q)\, ,\label{eqn:cla_eq_vel}
\end{equation}
where the velocity field is defined to satisfy 
\begin{equation}
    \dot{q}^i(t) = v^i (q(t),t) \, .
\end{equation}
This expression is compared to the equation obtained by stochastic variation.

\subsection{Evaluation of Stochastic Parallel Transport}

The stochastic parallel transport plays an important role in implementing the variation 
associated with stochastic variables, and we need to evaluate more quantitatively its behavior.
From the definition and the forward and backward SDE's, the stochastic parallel transports of the vielbein tensor are reexpressed as 
\begin{align}
\ud
\tensor{e}{_{\hat{a}}^i}
&= 
\ud t \left[
- U^j_\pm  {\Gamma^{i}}_{k j} 
\mp \nu g^{jl}  
(\partial_l {\Gamma^{i}}_{k j}   )
\pm  \nu
{\Gamma^{i}}_{\lambda j}
{\Gamma^{\lambda}}_{k l} 
 g^{jl}
\right]  
\tensor{e}{_{\hat{a}}^k}
- \sqrt{2\nu} {\Gamma^{i}}_{k j} 
\tensor{e}{_{\hat{a}}^k}
\tensor{e}{_{\hat{b}}^j}
\ud\widehat{B}^{\hat{b}}_\pm \, . \label{eqn:spt_rep}
\end{align}
The indices "$\pm$" correspond to the results of the forward and backward SDE's, respectively.

To find the stochastic parallel transport of the dual vielbein tensor $\ud \tensor{e}{_{\hat{a}}^i} := \ud \tensor{e}{_{\hat{a}}^i} (\widetilde{q}(t))$, we use the following relation: 
\begin{equation}
\tensor{e}{^{\hat{a}}_j} \circ \ud \tensor{e}{_{\hat{a}}^i}
=
- \tensor{e}{_{\hat{a}}^i} \circ \ud \tensor{e}{^{\hat{a}}_j}
\longrightarrow
 \tensor{e}{^{\hat{a}}_j} \ud \tensor{e}{_{\hat{a}}^i}
 = - \tensor{e}{_{\hat{a}}^i} \ud \tensor{e}{^{\hat{a}}_j} 
 - \ud \tensor{e}{^{\hat{a}}_j} \ud \tensor{e}{_{\hat{a}}^i} \, ,
 \label{eqn:ede+dee}
\end{equation}
which is obtained from the orthogonality condition
\begin{equation}
    \delta^i_j = \tensor{e}{_{\hat{a}}^i} \tensor{e}{^{\hat{a}}_j} \, .
\end{equation}
To calculate the second term on the right-hand side, we use Eq.\ (\ref{eqn:spt_rep}),   
\begin{align}
\ud \tensor{e}{^{\hat{a}}_j} \ud \tensor{e}{_{\hat{a}}^i}
&=
- \sqrt{2\nu} {\Gamma^{i}}_{k j} 
\tensor{e}{_{\hat{a}}^k}
\tensor{e}{_{\hat{b}}^j}
\ud\widehat{B}^{\hat{b}}_\pm \ud \tensor{e}{^{\hat{a}}_j} 
\nonumber \\
&=
- \sqrt{2\nu} {\Gamma^{i}}_{k j} 
\tensor{e}{_{\hat{b}}^j}
\ud\widehat{B}^{\hat{b}}_\pm 
(- \tensor{e}{^{\hat{a}}_j} \ud \tensor{e}{_{\hat{a}}^k} ) 
\nonumber \\
&=
- 2\nu |dt| g^{kl} 
{\Gamma^{i}}_{m k} 
{\Gamma^{m}}_{j l} 
 \, .
\end{align}
In the above expression, we maintain the lowest order term in $\ud t$.
These are applicable both in the forward and backward SDE's.

We further consider the forward and backward mean derivatives for an arbitrary vector $A^{i}(\widetilde{q}(t),t)$,
\begin{align}
\tensor{e}{_{\hat{a}}^i}
\uD_\pm A^{\hat{a}} 
&= \partial_t A^{i}
+
u^k_\pm \nabla_k A^{i} \pm \nu \Delta A^{i} \, . \label{eqn:formula_eda}
\end{align}
In this derivation, we used the Leibniz rule for the Stratonovich definition of the product, 
\begin{equation}
\ud A^{\hat{a}} = (\ud \tensor{e}{^{\hat{a}}_i}) \circ A^{i} + \tensor{e}{^{\hat{a}}_i} \circ (\ud A^{i}) \, .
\end{equation}

We now extend the formulation in Sec.\ \ref{sec:svm_particle} to the curved geometry with torsion. 
The stochastic Lagrangian, analogous to the classical Lagrangian in local coordinates, is given by
\begin{equation}
L_{sto} = \frac{\uM}{4} \eta_{\hat{a}\hat{b}} \left\{ 
\left(\uD_+ \widetilde{q}^{\hat{a}} \right) \left( \uD_+ \widetilde{q}^{\hat{b}} \right) + \left( \uD_- \widetilde{q}^{\hat{a}} \right) \left( \uD_- \widetilde{q}^{\hat{b}} \right) \right\} - V \left(\widetilde{q}\right)\, . \label{eqn:sto_lag_curve}
\end{equation}
This form represents the generalization of the flat-space Lagrangian that corresponds to the standard quantization (where $\alpha_A = 0$ and $\alpha_B = 1/2$, as introduced in Sec.\ \ref{sec:sto_action}). 
It was specifically selected so that, in the flat-space limit, the stochastic variation correctly reproduces the standard Schr\"{o}dinger equation. 
As discussed in Sec.\ \ref{sec:sto_action}, employing a more general quadratic form of the mean derivatives typically introduces viscous effects \cite{N_Koide_2012,V_Koide_2020}. 
How these quantum viscous effects intertwine with spacetime torsion is an intriguing open question, which we leave for future investigations.

Unlike the above deterministic case, 
the dynamics derived from this stochastic Lagrangian will depend on torsion. 
The mechanism for this coupling is the quantum fluctuations hidden in the mean derivatives $\uD_\pm$. These derivatives, through their connection to the Fokker-Planck equation, 
are coupled with vielbein and thus torsion.
Therefore, the stochastic optimization process introduces a coupling between the particle and torsion that has no classical analogue.
Substituting this into Eq.\ (\ref{eqn:sto_act}), we obtain the stochastic action. 
We then perform the stochastic variation of the action with respect to the local coordinates,
\begin{equation}
\widetilde{q}^{\hat{a}}(t) \longrightarrow \widetilde{q}^{\hat{a}}(t) + \delta f^{\hat{a}}(\widetilde{q}(t),t) \, ,
\end{equation}
where $\delta f^{\hat{a}}(x,t)$ is an arbitrary infinitesimal smooth function satisfying $\delta f^{\hat{a}}(x,t_i) = \delta f^{\hat{a}}(x,t_f)=0$. 
For example, the variation of the kinetic term is calculated as 
\begin{align}
\delta \int_{t_i}^{t_f} \ud t \uE \left[ 
\eta_{\hat{a}\hat{b}}  (D_\pm q^{\hat{a}})(D_\pm q^{\hat{b}})
\right]
&= 
2 \int_{t_i}^{t_f} dt \uE \left[ \eta_{\hat{a}\hat{b}}  
u_\pm^{\hat{a}} (q(t),t)  (D_\pm \delta q^{\hat{b}})
\right] \nonumber \\
&= 
- 2 \int_{t_i}^{t_f} \ud t  \uE \left[ 
 \delta q^{\hat{b}} 
g_{ij} (q,t)  e^j_{\hat{b}} (q,t)e^i_{\hat{a}}(q,t)
D_\mp u_\pm^{\hat{a}} (q(t),t)  
\right] \nonumber \\
&= 
- 2 \int_{t_i}^{t_f} \ud t   \uE \left[ 
g_{ik} \delta q^{k}   
\{
\partial_t u^i_\pm + u^j_\mp \nabla_j u^i_\pm \mp \nu \Delta u^i_\pm 
\}
\right] \, .
\end{align}
In the last line, we used Eq.\ (\ref{eqn:formula_eda}).
The stochastic variation of the stochastic action leads to 
the following equation: 
\begin{equation}
\partial_t (u^i_+ + u^i_-)
+
u^j_- \nabla_j u^i_+ +
u^j_+ \nabla_j u^i_-
- \nu \Delta (u^i_+ - u^i_-)  
= - \frac{2}{\uM}
g^{ij}\partial_j V
 \, . \label{eqn:vari_eq_1}
\end{equation}
By expressing $u^j_\pm$ in terms of $v^{i}$ and the gradient of $\ln \rho$, as given in Eqs.\ (\ref{eqn:u+torsion}) and (\ref{eqn:u-torsion}), Eq.\ (\ref{eqn:vari_eq_1}) can be recast into a more convenient form:
\begin{align}
& \partial_t v^i + v^j \nabla_j v^i + \frac{1}{\uM}
g^{ij}\partial_j V 
=
\nu^2  g^{ij}
\left\{
2\nabla_j \left(\frac{1}{\sqrt{\rho}}\Delta \sqrt{\rho} \right)
 -  R_{l j} g^{l m}\partial_m \ln \rho 
 \right\} 
+ \nu^2 \Sigma^i \, ,\label{eqn:general_vari_eq_torsion}
\end{align}
where $R_{ij} : ={R^k}_{ijk}$, and 
\begin{align}
\Sigma^i
&= 
 2 \ g^{ij}g^{lm}
\left\{
-(\partial_m \ln \rho) {S^n}_{jl} (\partial_n \ln \rho) 
+ {S^n}_{lj}  \nabla_n \partial_m \ln \rho  
- \nabla_m ({S^n}_{jl} \partial_n \ln \rho) 
\right\} \nonumber \\
& -  
K^j \nabla_j K^i + \Delta_{LB} K^i
+ g^{jm} (\partial_m \ln \rho) \nabla_j K^i 
+ K^j \nabla_j g^{ln}\partial_n \ln \rho \, . \label{eqn:general_sigma}
\end{align}
To simplify the numerous terms generated by this substitution, we utilized the generalized commutator of covariant derivatives for an arbitrary vector $A^i$,
\begin{align}
[ \nabla_j,  \nabla_k] A^i 
&= {R^i}_{l j k} A^l 
- 2 {S^l}_{k j}  \nabla_l A^i \, ,
\end{align}
along with the following useful identities:
\begin{align}
g^{kl} (\partial_l \ln \rho) \nabla_k g^{ij}(\partial_j \ln \rho) 
%&= g^{kl}g^{ij} (\partial_l \ln \rho) \{ \nabla_j \partial_k \ln \rho - 2 {S^\rho}_{jk} \partial_\rho \ln \rho \} \nonumber \\
&= \frac{1}{2} g^{ij} \nabla_j (g^{kl} (\partial_l \ln \rho) (\partial_k \ln \rho)) \nonumber \\
&- 2 g^{kl}g^{ij} (\partial_l \ln \rho) {S^\rho}_{jk} \partial_\rho \ln \rho \, , \\
g^{km}g^{ij}\nabla_k \nabla_m \partial_j \ln \rho 
&= g^{ij}\nabla_j \Delta_{LB} \ln \rho -g^{ij} R_{k j} g^{kl}\partial_l \ln \rho \nonumber \\
& + 2 g^{ij}  g^{kl} ( {S^m}_{kj}  \nabla_m \partial_l \ln \rho  -\nabla_l ({S^m}_{jk} \partial_m \ln \rho) ) \, , \\
\frac{1}{\sqrt{\rho}} g^{ij} \nabla_i \partial_j \sqrt{\rho}
&=
\frac{1}{2} g^{ij} 
\left(
\nabla_i \partial_j \ln \rho + \frac{1}{2}(\partial_j \ln \rho) (\partial_i \ln \rho)
\right) \, .
\end{align}
Note that we have not yet used the total antisymmetry for the torsion tensor in Eq.\ (\ref{eqn:general_vari_eq_torsion}).

All terms on the right-hand side in Eq.\ (\ref{eqn:general_vari_eq_torsion}) stem from fluctuation (quantum effects) and hence, 
in the vanishing limit of fluctuations $\nu \rightarrow 0$, the optimized equation in the stochastic variation reproduces that in the classical variation (\ref{eqn:cla_eq_vel}). 
The first term on the right-hand side corresponds
to the gradient of Bohm’s quantum potential, while the
second reflects quantum fluctuations and spatial curvature,
as shown in Ref.\ \cite{N_Koide_2019}. 
The newly introduced term $\Sigma^i$ captures the interaction between quantum fluctuations
and torsion.

Before considering the totally antisymmetric torsion tensor, 
we investigate the momentum of the free particle $V=0$, 
\begin{equation}
 \uE\left[ \uM v^{i} (\widetilde{q}(t),t)\right] = \uM \int \ud^D q \, \rho(q,t) v^{i} (q,t) \, .
\end{equation}
This time evolution is calculated through Eqs.\ (\ref{eqn:eofconti}) and (\ref{eqn:general_vari_eq_torsion}),
\begin{equation}
    \partial_t \left( \rho v^{i} \right)
 = -  \nabla_j \left\{ 
 \rho \left( v^{i} v^{j} -\nu^2 g^{il}  g^{jm}  \nabla_l \partial_m \ln \rho \right) \right\}
 + \sigma^i \, ,
\end{equation} 
where
\begin{equation}
\sigma^i = 
-\nu^2 \rho 
\left\{
- K^j \nabla_j K^i + g^{jl}\nabla_j \nabla_l K^i
+   g^{jl}\partial_l \ln \rho \nabla_j K^i + 
 g^{ij} K^l \nabla_l \partial_j \ln \rho 
\right\} \, .
\end{equation}
Therefore, the conservation of momentum for a free particle requires that the source term $\sigma^i$ vanishes. 
A sufficient condition for this is $K^i=0$, which is naturally satisfied for a totally antisymmetric torsion tensor.

\subsection{Totally Antisymmetric Torsion}

While the properties of the torsion tensor have not been explicitly used thus far, we can simplify Eq.\ (\ref{eqn:general_vari_eq_torsion}) by using its total antisymmetry. 
In a 3-dimensional space, this constraint aligns torsion with the
homogeneous and isotropic properties of the space, expressed as 
\begin{equation}
S_{abc} = s(t)\epsilon_{abc} \, ,   
\end{equation}
where $s(t)$ is a scalar field and $\epsilon_{abc}$
denotes the 3-dimensional Levi-Civita tensor, which is defined by multiplying $\sqrt{{\rm det}\, g}$ with the Levi-Civita symbol. 
This follows directly from the Hodge duality relation for $D=3$,
\begin{equation}
\tensor{\star S}{_{a_1 \ldots a_{D-3}}} = \frac{1}{3!}\tensor{\epsilon}{^{b_1 b_2 b_3}_{a_1 \ldots a_{D-3}}}\tensor{S}{_{b_1 b_2 b_3}} \, .
\end{equation}
Here, the Hodge star is given by the scalar field $s(t)$.
In this formulation, torsion
satisfies $S_i = 0$ and becomes equivalent to the contortion
tensor;
\begin{equation}
K_{ijk} = S_{ijk} \, .
\end{equation}
%This choice of torsion is also 
%preferred from a metric-affine gravity standpoint as it allows for a
%continuous limit of vanishing torsion, thereby recovering GR \cite{C_Fabbri_2014}.
The totally antisymmetric torsion tensor satisfies 
\begin{equation}
{S^{l}}_{k j}{S^{k i}}_{l} = 2 s^2\delta^i_j \, 
\end{equation}

For this idealized situation, $S^i = K^i = 0$ and thus our velocity equation (\ref{eqn:general_vari_eq_torsion}) is simplified as 
\begin{align}
&    \partial_{t}v^i + v^j \nabla_j v^i 
    + \frac{1}{\uM}g^{i j}\partial_j V   
    - \nu^2 g^{i j} 2\nabla_j \left(\frac{1}{\sqrt{\rho}}\Delta\sqrt{\rho}\right) 
    \nonumber \\
&    =- \nu^2 g^{i j}\left[
    R_{k j}
    g^{k l}\partial_{l}\ln\rho 
    + 2 (\partial_k \ln\rho) \nabla_l {S^{l k}}_{j} \right]  \, .
\end{align}
To separate the Levi-Civita and Torsional contributions hidden in $R_{k j}$ and $\nabla_i$, 
we finally obtain 
\begin{align}
&    \partial_{t}v^i + v^k \mathring{\nabla}_k v^i 
   + \frac{1}{\uM}g^{ik}\partial_k V 
   -2 \nu^2 g^{ik}
    \partial_k \left(\frac{1}{\sqrt{\rho}}\mathring{\Delta}\sqrt{\rho}\right)  \nonumber \\
&= 
    - \nu^2 g^{ik}\biggl\{ 2 (\partial_k \ln\rho)s^2 + \mathring{R}_{l k}
    g^{l j}\partial_{j}\ln\rho 
    + (\partial_j \ln\rho) \mathring{\nabla}_l {S^{l j}}_{k} 
    \biggr\} \nonumber \\
    &= 
    - \nu^2 g^{ik}\biggl\{ 2 (\partial_k \ln\rho)s^2 + \mathring{R}_{l k}
    g^{l j}\partial_{j}\ln\rho \biggr\}
    - \nu^2  (\partial_j \ln\rho) g^{jm} (\mathring{\nabla}_l {S^{l}}_{mk}) g^{ki} 
    \, , \label{eqn:v_tastorsion}
\end{align}
where $\mathring{\Delta} := g^{ij} \mathring{\nabla}_i \mathring{\nabla}_j$.  
In the derivation, we used 
\begin{equation}
(\partial_j \ln\rho) \nabla_l {S^{l j}}_{k} = (\partial_j \ln\rho)\mathring{\nabla}_l {S^{l j}}_{k}\, ,
\end{equation}
and the Riemann-Cartan tensor splits into its Levi-Civita tensor part ($\mathring{R}_{ij}$)
and torsional contributions,
\begin{equation}
    R_{ij} = \mathring{R}_{ij} - \mathring{\nabla}_k {S^{k}}_{ij}
    + {S^{k}}_{l j}{S^{l}}_{i k} \, .
\end{equation}

%It is easy to see that Eq.\ (\ref{eqn:v_tastorsion}) reduces to the classical equation (\ref{eqn:cla_eq_vel}) in the vanishing limit of $\nu$.
%That is, our extended formulation of SVM is yet a natural generalization of the classical variational method in curved geometry with torsion.

\subsubsection{Homogeneous Curved Geometry without Torsion}

We consider quantum mechanics on a homogeneous curved geometry where the line element is defined by 
\begin{equation}
    \ud s^2 = a^2 \left\{ \frac{1}{1-Kr^2} (\ud r)^2 + r^2 (\ud \theta)^2 + r^2 \sin^2 \theta (\ud \phi)^2 \right\} \, ,
\end{equation}
where $a$ is a positive real constant and $K$ takes the value $1$ (spherical), $0$ (Euclidean) or $-1$ (hyperspherical), 
as a simplified case of the Friedmann–Lema\^{i}tre–Robertson–Walker (FLRW) metric in cosmology.

Substituting this into Eq.\ (\ref{eqn:v_tastorsion}), the equation for the velocity field becomes 
\begin{align}
&    \partial_{t}v^i + v^k \mathring{\nabla}_k v^i 
   + \frac{1}{\uM}g^{ik}\partial_k V 
   -2 \nu^2 g^{ik}
    \partial_k \left(\frac{1}{\sqrt{\rho}}\mathring{\Delta}\sqrt{\rho}\right)  \nonumber \\
    &= 
    - \nu^2 g^{ij} \partial_j \biggl\{ 2 \ln\rho\, s^2 
    - \frac{1}{3} \mathring{R} \ln\rho \biggr\}
    \, , \label{eqn:v_eq_homogeneous}
\end{align}
where $\mathring{R}$ is the scalar curvature calculated from the Levi-Civita connection. 
In this derivation, we used 
\begin{align}
R_{ij} &= -\frac{2K}{a^2}g_{ij} = - \frac{1}{3} \mathring{R} g_{ij}\, ,\\
(\mathring{\nabla}_l {S^{l}}_{ij}) &= 0\, .
\end{align}
The coupled dynamics composed of the equation of continuity of $\rho$ (\ref{eqn:eofconti}) and the above optimized equation (\ref{eqn:v_eq_homogeneous}) describes 
quantum dynamics of a particle of mass $\uM$ under the homogeneous curved geometry with torsion. 
To simplify the equations, we introduce 
the wave function 
\begin{equation}
    \Psi = \sqrt{\rho} e^{\ii \theta} \, ,
\end{equation}
where the phase $\theta$ is given by the velocity potential, $v^i= 2 \nu g^{ij} \partial_j \theta$. 
The dynamics of $\theta$ is obtained from Eq.\ (\ref{eqn:v_eq_homogeneous}).
%\begin{align}
%\partial_t \theta + \nu g^{ij} (\partial_i \theta) (\partial_j \theta) 
%+ \frac{1}{2\nu \uM} V - \nu \frac{1}{\sqrt{\rho}}\mathring{\Delta}_{LB}\sqrt{\rho} = 
%- \frac{\nu}{2} \biggl\{ 2 s^2 \ln\rho
%    - \frac{1}{3} \mathring{R} \ln\rho \biggr\}\, .
%\end{align}
In short, by choosing $\nu=\hbar/(2m)$, the non-linear Schr\"{o}dinger equation is obtained, 
\begin{align}
    \ii \hbar\partial_t \Psi 
    = \left[ -\frac{\hbar^2}{2\uM} \Delta  +  V
    -
    \frac{\hbar^2}{2\uM} \left(
    \frac{\mathring{R}}{6} - s^2
    \right) \ln |\Psi|^2  \right]\Psi \, . \label{eqn:log-schriedinger}
\end{align}
Here we used
\begin{align}
    \Delta \Psi 
     &=
     \frac{1}{\sqrt{\rho}} \Delta_{LB} \sqrt{\rho}
     -g^{ij}(\partial_i \theta)(\partial_j \theta)
     + \ii g^{ij}
     \left(
     \partial_i \partial_j \theta - 
     \mathring{\Gamma^{m}}_{j i}
     \partial_m \theta +  (\partial_j \theta) (\partial_i \ln \rho) 
     \right) \, ,\\
 \frac{1}{\sqrt{\rho}}\Delta \sqrt{\rho}   
 &= g^{ij} \left( \frac{\partial_i \partial_j \rho}{2\rho} 
 - \frac{(\partial_i \rho)(\partial_j \rho)}{4\rho^2} \right)
 - \frac{1}{2}g^{i j} 
 \mathring{\Gamma^{k}}_{j i}
 \partial_k \ln \rho  \, .
\end{align}
It is easy to see that this reduces to the standard Schr\"{o}dinger equation in the flat space without torsion. 
This result indicates that the wave function formalism can be consistently extended to a homogeneous curved geometry with totally antisymmetric torsion. 
However, the equation acquires non-linear terms, which arise from the couplings between quantum fluctuations and spatial curvature, and between quantum fluctuations and torsion.
Similar non-linear equations for $s=0$ and for $\mathring{R}$ are obtained in Refs.\ \cite{N_Koide_2019} 
and \cite{Armin_Koide_2025}, respectively.

Non-linear modifications to the Schr\"{o}dinger equation have been extensively studied in various contexts, including soliton dynamics, wave function collapse models, and alternative formulations of quantum mechanics \cite{kostin,Bialynicki75,Bialynicki76,kibble,dekker,DIOSI1987377,PhysRevA.40.1165,PhysRevLett.62.485,WEINBERG1989336,PhysRevA.40.3387,Polchin,penrorse,CZACHOR19971,PhysRevA.57.4122,RevModPhys.85.471,CZACHOR2002139,N_Koide_2019,G_Koide_2018,dieter,al}.
Crucially, the phenomenological extension of quantum mechanics proposed by Bialynicki-Birula and Mycielski \cite{Bialynicki75,Bialynicki76} identified the logarithmic non-linear term as the most plausible form that rigorously preserves essential physical properties, such as the continuity equation (conservation of probability), the separability of non-interacting subsystems, the Born rule, and the Ehrenfest theorem. 
It is of fundamental importance that our derived Eq.\ (\ref{eqn:log-schriedinger}) takes exactly this theoretically favored logarithmic form, meaning it inherently satisfies all these stringent physical requirements. 
Our primary novelty lies in the physical origin of this equation. In contrast to Refs.\ \cite{Bialynicki75,Bialynicki76}, where the coefficient of the logarithmic term is introduced merely as a phenomenological free parameter, our framework derives this exact non-linear term rigorously from first principles (SVM). Furthermore, we explicitly determine its coefficient geometrically, revealing it to be governed by the competitive interplay between spatial curvature and torsion.
It is also interesting to mention that 
similar logarithmic non-linear terms can emerge even in the phenomenological quantization of classical dissipative systems, although they often feature purely imaginary coefficients \cite{kostin,dekker,PhysRevA.40.3387,G_Koide_2018,dieter}.

Furthermore, it gives rise to other well-studied theoretical implications, including: (i) asymptotic classicality, where initially non-orthogonal states can evolve to become perfectly distinguishable \cite{Mielnik_1980}; (ii) the potential for superluminal signaling \cite{gisin_1989}; and (iii) the violation of the mixture equivalence principle, wherein distinct physical preparations of the same density matrix evolve differently \cite{pttr-6kj7}. Crucially, since the strength of our non-linear term is governed by the torsion magnitude, the experimental non-observation of these phenomena provides a powerful method to constrain its value, implying that the magnitude of $s(t)$ must be extremely small.

As discussed in Refs.\ \cite{Bialynicki75,Bialynicki76}, 
the positive sign of our non-linear term is required to have a stationary solution 
of the non-linear Schr\"{o}dinger equation. 
In our equation, the sign of the log-non-linear term depends on 
the relative difference of magnitude of $s^2$ and $\mathring{R}$. 
Therefore, for a stationary solution to exist, the magnitude of torsion must satisfy the following inequality,
\begin{equation}
s^2 \le \frac{\mathring{R}}{6} \, .
\end{equation}
The Ricci scalar is given by $\mathring{R} = 6K/a^2$. From the Planck 2018 \cite{Planck2018} data we employ
\begin{equation}
    \Omega_K \coloneqq -\frac{K}{(a_0 H_0)^2} \approx 0.0007 \, ,
\end{equation}
which, with normalization $a_0 = 1$, yields $K = -0.0007 H_0^2$. Choosing the Planck value $H_0 = 67.4 \text{km}/(\text{s}\cdot\text{Mpc}) \approx 1.438\times10^{-39} \text{MeV}$ we then get
\begin{equation}
    K \approx -1.45 \times 10^{-81} \text{MeV}^2 \, .
\end{equation}
Thus, the above restriction reduces to
\begin{equation}
s^2 \le -1.45 \times 10^{-81}\frac{1}{a^2} \text{MeV}^2 \, .
\end{equation}
Since the squared torsion $s^2$ and $a^2$ must be non-negative, this condition cannot be satisfied. Consequently, these observations imply that our non-linear Schr\"{o}dinger equation does not admit stationary solutions under the current cosmological constraints.
We should however add a cautionary remark regarding the application of these cosmological constraints. 
Since our current theoretical framework is non-relativistic and incorporates only spatial geometry, directly applying observational bounds derived from a fully relativistic four-dimensional cosmological framework is not strictly exact. 
Therefore, the constraint based on the cosmological data should be understood merely as an order-of-magnitude estimate. A more rigorous evaluation would require a fully covariant extension of SVM, which remains an important subject for future research.

An additional constraint on the torsion magnitude arises from the fact that torsion could induce small shifts in the hydrogen atom’s energy levels. While these shifts are likely negligible compared to the Coulomb potential, evaluating the potential energies at the Bohr radius 
$R_B$ (assuming $\ln |\Psi (R_B)|^2 \sim 1$) yields the following inequality:
\begin{equation}
\sqrt{\left|\frac{\mathring{R}}{6} -s^2 \right|} << \sqrt{
2\alpha
\frac{m_e}{\hbar^2 R^2_B}
}
\approx 5.3 \times 10^{-3} \, {\rm MeV} \sim 10^{-2} m_e
\end{equation}
where $\alpha$ is the fine-structure constant and $m_e$ is the electron
mass. 
These derived bounds are consistent with the significantly more stringent experimental constraint reported in Ref.\ \cite{PhysRevLett.100.111102}, which suggests $s \le 10^{-31}$ GeV under the assumption of Lorentz symmetry violation.

Torsion has been shown to induce non-linearity in the
Dirac equation \cite{Hehl1,zecca,POPLAWSKI201073,POPLAWSKI2013575,Fabbri}. 
This non-linearity, however, is
fundamentally different from the type examined here. As
demonstrated in Ref.\ \cite{PhysRevD.98.104027}, the non-linearity in the Dirac
equation can, in turn, give rise to the non-linearity in the
Schr\"{o}dinger-Newton equation in its non-relativistic limit.
This non-linearity, however, arises from the gravitational
potential rather than torsion, highlighting the distinct
mechanisms driving the two effects.

In the SVM quantization, both the form and the coefficient
of the non-linear term depend on the choice of
Lagrangian. A time-dependent Lagrangian that violates
time-reversal symmetry gives rise to a different non-linear
term with a purely imaginary coefficient \cite{G_Koide_2018,PhysRevA.40.3387}.
Meanwhile, choosing a hyperspherical surface yields a
log-non-linear term with a negative real coefficient associated
with the Ricci scalar \cite{N_Koide_2019}. Notably, the Gross-
Pitaevskii equation is obtained by the SVM quantization
of an ideal fluid Lagrangian \cite{N_Koide_2012}. These examples underscore
the adaptability of SVM as a natural framework for
extending quantum theory non-linearly.

Finally, we remark on the implications for the geometrical trinity of gravity, at least, for torsion. 
Classically, this principle asserts the equivalence of gravity descriptions based on curvature, torsion, or non-metricity \cite{beltran2019}. 
However, our results indicate that, within the framework of SVM quantization this equivalence may be broken at the quantum level.
The non-linear term in Eq.\ (\ref{eqn:log-schriedinger}), which originates purely from quantum fluctuations, depends on the specific combination $\mathring{R}/6 - s^2$. If the Trinity held strictly, one should be able to reproduce the effects of any Levi-Civita curvature $\mathring{R}$ solely through torsion $s$ in a flat background. 
This would require a real solution for $s$ satisfying $\mathring{R} = 6s^2$. 
However, for spaces with negative scalar curvature ($\hat{R} < 0$), no such real solution exists. 
 This suggests that quantum fluctuations distinguish between the geometric effects of curvature and torsion through their interaction with matter. Thus, the violation of the geometrical trinity at the quantum level serves as a reminder that the classical equivalence of these geometric descriptions does not guarantee their physical equivalence upon quantization, showing the inherent difficulty in the unification of GR and quantum mechanics.

%It is important to emphasize that the introduction of 
%the non-linear Schr\"{o}dinger equation is a consequence of our idealization of 
%a homogeneous geometry and a totally antisymmetric torsion. 
%In a more general scenario, where the geometry is inhomogeneous or the torsion has a more complex structure, it is not guaranteed that quantum dynamics can be expressed in terms of a wave function. 
%In such cases, the fundamental description of the system's time evolution is given by the coupled equations: 
%the stochastic equation of motion for the velocity field (\ref{eqn:general_vari_eq_torsion}), and the continuity equation for the probability distribution (\ref{eqn:eoc}). 
%The state of the system would then be determined by solving these two real-valued equations directly.

\section{Information Geometry} \label{sec:inf_geo}

While the present work has focused exclusively on the role of torsion, the geometrical landscape of MAG offers a complementary perspective through the concept of non-metricity. 
Interestingly, the gravitational interaction can be equivalently described by non-metricity in a flat, torsion-free spacetime. Historically, this approach was formalized as Symmetric Teleparallel Gravity by Nester and Yo \cite{nester1999}, who demonstrated that non-metricity can mediate the gravitational force in a geometry devoid of curvature and torsion. 
Recently, this framework has experienced a resurgence through its non-linear extension, $f(Q)$ gravity \cite{beltran2019,heisenberg2024}.
In this context, we turn our attention to another framework dealing with non-metricity: information geometry \cite{amari2000,amari1985,Ay2017}.

Information geometry treats a family of probability distributions as a statistical manifold. In this framework, a quantity known as "divergence" acts as a distance, yet it is distinguished from a standard metric distance by its lack of symmetry. 
This asymmetry gives rise to a rich geometrical structure characterized by dual connections, which inherently violate metric compatibility (i.e., they generate non-metricity). 
Although the application of information geometry to generalized gravity theories has been initiated by Iosifidis and Pallikaris \cite{iosifidis2023}, 
it remains to be fully explored. 
To motivate a new pathway for incorporating non-metricity, we briefly summarize the formulation of information geometry and highlight its structural similarity to the SVM framework discussed thus far.

\subsection{Statistical Manifold}

A family of probability distributions $p(x;\theta)$ parametrized by $\theta = (\theta^1, \dots, \theta^n)$ forms a statistical manifold, 
where each point $\theta$ corresponds to a single distribution. 
The Fisher information matrix $g_{ij}(\theta)$ is considered to be a natural choice for the metric of the statistical manifold \cite{chentsov1982}, 
\begin{equation}
    g_{ij}(\theta) = E_{\theta}\left[ \left(\frac{\partial}{\partial\theta^i} \log p(x;\theta)\right) \left(\frac{\partial}{\partial\theta^j} \log p(x;\theta)\right) \right] \, ,
\end{equation}
where $E_{\theta}[\cdot]$ is the expectation value with $p(x;\theta)$.

In information geometry, we can define a continuous family of affine connections, known as $(\alpha)$-connections, 
\begin{equation}
    {}^{(\alpha)}{\Gamma^{i}}_{jk}(\theta) = \tensor{\mathring{\Gamma}}{^{i}_{jk}}(\theta) - \frac{\alpha}{2}g^{il}T_{jkl}(\theta) \, ,
\end{equation}
where $\tensor{\mathring{\Gamma}}{^{i}_{jk}}$ are the Christoffel symbols of the Levi-Civita connection and $\alpha$ is any real constant.
The third rank tensor $T_{ijk}(\theta)$ is called the cubic tensor which characterizes the non-metricity of the statistical manifold and is defined by 
\begin{equation}
    T_{ijk}(\theta) = E_{\theta}\left[ \left(\frac{\partial \log p(x;\theta)}{\partial\theta_i}\right) \left(\frac{\partial \log p(x;\theta)}{\partial\theta_j}\right) \left(\frac{\partial \log p(x;\theta)}{\partial\theta_k}\right) \right] \, .
\end{equation}
In standard information geometry, these connections are assumed to be torsion-free, meaning their connection coefficients are symmetric in the lower indices.

In this geometric framework, tangent vectors are represented as directional derivative operators. 
In the local coordinate system $\{\theta \}$, the set of partial derivative operators $\{\partial_i := \partial/\partial\theta^i\}$ forms a basis for the tangent space at each point. 
Any tangent vector field $X$ can therefore be expressed as a linear combination of these basis vectors: 
$X = X^i \partial_i$, where $X^i$ are the component functions.

Each $(\alpha)$-connection defines a notion of parallel transport and differentiation on the manifold. 
The covariant derivative of a vector field $V$ along another vector field $Z$, denoted as ${}^{(\alpha)}\nabla_Z V$, is given in local coordinates by
\begin{equation}
    ({}^{(\alpha)}\nabla_k V)^i := \partial_k V^i + {}^{(\alpha)}\tensor{\Gamma}{^i_{lk}} V^l \, .
\end{equation}
A crucial feature of this structure is the concept of duality. The ${}^{(-\alpha)}\nabla$ connection is said to be dual to the ${}^{(\alpha)}\nabla$ connection with respect to the Fisher metric. 
This duality is expressed by a generalized product rule that relates the directional derivative of an inner product to the two connections. For any vector fields $X, Y, Z$:
\begin{equation}
    Z\langle X,Y\rangle 
    = \langle {}^{(\alpha)}\nabla_Z X,Y\rangle + \langle X, {}^{(-\alpha)}\nabla_Z Y\rangle \, , \label{eqn:non-metricity}
\end{equation}
where $\langle X,Y\rangle := g_{ij} X^i Y^j$ is the inner product induced by the Fisher metric, and $Z\langle X,Y\rangle$ is the directional derivative of this scalar function along $Z$. 
This relation can be seen as a generalization of the product rule for derivatives in Euclidean space. 
Physically, this relation signifies that the $(\alpha)$-connection is not metric-compatible, meaning the length of a vector is generally not preserved under parallel transport using only an $(\alpha)$-connection.

%\subsection{Divergence}

A divergence $D(\theta_p\|\theta_q)$ is a function that quantifies the difference between two distributions with $\theta_p$ and $\theta_q$. While not a true distance, it must satisfy several properties analogous to distance:
\begin{enumerate}
    \item $D(\theta_p\|\theta_q) \ge 0$ for all points $\theta_p, \theta_q$ on the manifold. 
    \item $D(\theta_p\|\theta_q) = 0$ if and only if $\theta_p=\theta_q$. 
    \item For two infinitesimally close distributions with parameters $\theta$ and $\theta+\mathrm{d}\theta$, the divergence is related to the Fisher metric as:
    \begin{equation}
        D(\theta+\mathrm{d}\theta\|\theta) = \frac{1}{2}g_{ij}(\theta)\mathrm{d}\theta^i \mathrm{d}\theta^j + O(\|\mathrm{d}\theta\|^3) \, .
    \end{equation}
\end{enumerate}

From a given divergence function, 
one can recover the metric and the dual connections for $\alpha=1$ (e-connection) and $\alpha=-1$ (m-connection) through differentiation:
\begin{align}
    g_{ij}(\theta) 
    &= - \left[ \frac{\partial}{\partial(\theta_p)^i} \frac{\partial}{\partial(\theta_q)^j} D(\theta_p\|\theta_q) \right]_{\theta_p=\theta_q=\theta}\, , \\
    ^{(e)}\Gamma_{kij}(\theta) 
    &:= g_{kl} (\theta)\, ^{(e)}{\Gamma^l}_{ij}(\theta) 
    = - \left[ \frac{\partial}{\partial(\theta_p)^i} \frac{\partial}{\partial(\theta_p)^j} \frac{\partial}{\partial(\theta_q)^k} D(\theta_p\|\theta_q) \right]_{\theta_p=\theta_q=\theta} \label{eqn:e-conn} \, , \\
    ^{(m)}\Gamma_{kij}(\theta) 
    &:= g_{kl} (\theta)\, ^{(m)}{\Gamma^l}_{ij}(\theta) 
    = - \left[ \frac{\partial}{\partial(\theta_q)^i} \frac{\partial}{\partial(\theta_q)^j} \frac{\partial}{\partial(\theta_p)^k} D(\theta_p\|\theta_q) \right]_{\theta_p=\theta_q=\theta} \, .
\end{align}

\subsection{The Exponential Family}
\label{sec:ef}

An exponential family is a set of probability distributions that can be written in the form:
\begin{equation}
    p(x;\theta) = \exp\left((\theta)^i F_{i}(x) - \psi(\theta)\right) \, ,
\end{equation}
where $\theta$ are the natural parameters and $\psi(\theta)$ is a scalar function called the potential. For this family, the Fisher metric $g_{ij}$ and the cubic tensor $T_{ijk}$ are given by the second and third derivatives of this potential:
\begin{align}
    g_{ij}(\theta) &= \frac{\partial^2 \psi(\theta)}{\partial\theta^i \partial\theta^j} \label{eq:g_from_psi} \\
    T_{ijk}(\theta) &= \frac{\partial^3 \psi(\theta)}{\partial\theta^i \partial\theta^j \partial\theta^k} \, .
\end{align}
Equation \eqref{eq:g_from_psi} proves that $\psi(\theta)$ is a convex function, 
as its Hessian matrix, the Fisher metric, is positive semi-definite by definition.

The convexity of $\psi(\theta)$ allows us to define a Legendre transformation, which introduces a dual coordinate system $\eta$, the expectation parameters:
\begin{equation}
    \eta_{i} = \frac{\partial\psi(\theta)}{\partial\theta^{i}} \, .
\end{equation}
This transformation also defines a dual potential $\phi(\eta)$ via the Legendre transformation:
\begin{equation}
    \phi(\eta) = \max_{\theta} \{ (\theta)^i \eta_i - \psi(\theta) \} \, .
\end{equation}
The pair of dual coordinates $(\theta, \eta)$ is a central feature of the exponential family.

Furthermore, the potential $\psi(\theta)$ naturally defines the Bregman divergence:
\begin{equation}
    D_{BR}(\theta_p\|\theta_q) := \psi(\theta_p) - \psi(\theta_q) - (\theta_p - \theta_q)^i \frac{\partial\psi(\theta_q)}{\partial(\theta_q)^i} \, .
\end{equation}
Using this divergence, we can explicitly calculate the e-connection coefficients from the general formula in Eq.\ (\ref{eqn:e-conn}). Differentiating $D_{BR}$ with respect to its arguments yields:
\begin{equation}
    ^{(e)}\Gamma_{kij}(\theta) = 0 \, .
%    ^{(e)}\Gamma_{kij}(\theta) = - \left[ \frac{\partial}{\partial(\theta_p)^i} \frac{\partial}{\partial(\theta_p)^j} \frac{\partial}{\partial(\theta_q)^k} D(\theta_p\|\theta_q) \right]_{\theta_p=\theta_q=\theta} = 0 \, .
\end{equation}
Since the e-connection coefficients vanish in the $\theta$-coordinates, the corresponding geodesics (e-geodesics) are straight lines in this coordinate system.

Dually, we can define the dual Bregman divergence, $D_{BR}^*$, based on the dual potential $\phi(\eta)$:
\begin{equation}
    D_{BR}^*(\eta_p\|\eta_q) := \phi(\eta_p) - \phi(\eta_q) - (\eta_p - \eta_q)_i \frac{\partial\phi(\eta_q)}{\partial(\eta_q)_i} \, .
\end{equation}
The m-connection coefficients are then defined analogously using this dual divergence:
\begin{equation}
    ^{(m)}\tilde{\Gamma}^{ijk}(\eta) = - \left[ \frac{\partial}{\partial(\eta_p)_i} \frac{\partial}{\partial(\eta_p)_j} \frac{\partial}{\partial(\eta_q)_k} D_{BR}^*(\eta_p\|\eta_q) \right]_{\eta_p=\eta_q=\eta} = 0 \, .
\end{equation}
A calculation parallel to that for the e-connection confirms that these coefficients also vanish. 
This proves that m-geodesics are straight lines in the $\eta$-coordinate system. 
A manifold exhibiting this property is said to possess a dually flat structure.
Note that the $(\alpha)$-connection in natural parameters corresponds to the $(-\alpha)$-connection in dual coordinates, ie. the expectation parameters.

The physical significance of the $(\alpha)$-connection in information geometry has not yet been fully elucidated. Nevertheless, for the Gaussian statistical manifold, the m-connection has been identified with the Brownian bridge, representing a perfectly random motion \cite{koide2026}. 
See also Refs.\ \cite{fujiwara1995,ohara2009} regarding the interpretation of deviations from the m-connection.

\subsection{Euclidean Limit}
\label{sec:diver}

We demonstrate how this generalized geometric structure contains the familiar Euclidean distance as a special case \cite{amari2000}. This occurs when the manifold is self-dual, which corresponds to the condition that the cubic tensor vanishes, i.e., $T_{ijk}=0$.

The vanishing of the cubic tensor implies that the potential $\psi(\theta)$ must be a polynomial of at most degree 2 in $\theta$. 
Consequently, the metric components, $g_{ij}(\theta) = \partial_i \partial_j \psi(\theta)$, must be constant. 
Ignoring irrelevant constant and linear terms, the potential can be written as a simple quadratic form:
\begin{equation}
    \psi(\theta) = \frac{1}{2}g_{kl}\theta^k \theta^l \, .
\end{equation}
Under this condition, the dual coordinates $\eta_i$ become linearly related to the $\theta$ coordinates:
\begin{equation}
    \eta_i = \frac{\partial \psi}{\partial \theta^i} = g_{ik} \theta^k \, .
\end{equation}
Let us now evaluate the Bregman divergence for this self-dual case. Substituting the quadratic potential into the definition of $D_{BR}(\theta_q\|\theta_p)$ gives:
\begin{align}
    D_{BR}(\theta_q\|\theta_p) 
%    &= \psi(\theta_q) - \psi(\theta_p) - ((\theta_q)^i - (\theta_p)^i)(\eta_p)_i \\
%    &= \frac{1}{2}g_{ij}(\theta_q)^i (\theta_q)^j - \frac{1}{2}g_{ij}(\theta_p)^i (\theta_p)^j - ((\theta_q)^i - (\theta_p)^i)g_{ik}(\theta_p)^k \nonumber \\
    &= \frac{1}{2}g_{ij}\left( (\theta_q)^i - (\theta_p)^i \right) \left( (\theta_q)^j - (\theta_p)^j \right) \, .
\end{align}
This explicit recovery of Euclidean geometry is a foundational result in information geometry, providing a strong justification that divergence can be considered a natural generalization of distance \cite{amari2000}.

\subsection{Generalized Pythagorean Theorem}
\label{sec:pythagoras}

The Bregman divergence, defined on a dually flat manifold, satisfies a generalized Pythagorean theorem \cite{amari2000}. 

Let $\theta_p, \theta_q, \theta_r$ be three points on the manifold. The theorem states that if the m-geodesic from $\theta_p$ to $\theta_q$ is orthogonal at $\theta_q$ to the e-geodesic from $\theta_q$ to $\theta_r$, there exists the following relation:
\begin{equation}
    D_{BR}(\theta_p \| \theta_r) = D_{BR}(\theta_p \| \theta_q) + D_{BR}(\theta_q \| \theta_r) \, .
    \label{eq:pythagoras_m_e}
\end{equation}
This is analogous to the standard Pythagorean theorem for a right triangle, where $D_{BR}$ plays the role of the squared distance.

Dually, if the e-geodesic from $\theta_p$ to $\theta_q$ is orthogonal at $\theta_q$ to the m-geodesic from $\theta_q$ to $\theta_r$, the dual Pythagorean theorem holds:
\begin{equation}
    D_{BR}(\theta_r \| \theta_p) = D_{BR}(\theta_r \| \theta_q) + D_{BR}(\theta_q \| \theta_p) \, .
    \label{eq:pythagoras_e_m}
\end{equation}
Note that the arguments of the divergence are reversed, $D_{BR}(\theta_r \| \theta_p)$ which corresponds to the dual Bregman divergence $D_{BR}^* (\theta_p \| \theta_r)$ in the $\eta$-coordinates.

\subsection{Gravitational Aspect of Information}

As mentioned earlier, m-geodesics can be understood as representing purely random motion \cite{koide2026}.
This correspondence offers a powerful interpretation akin to Einstein's equivalence principle. In general relativity, a free particle follows a geodesic in curved spacetime, appearing as a straight line in a local free-falling frame. This leads to an Equivalence Principle for Information: on any statistical manifold, an unbiased random process follows a geodesic. This principle redefines randomness not as noise, but as free motion guided by information geometry.
Table \ref{tab:analogy} summarizes the analogy between general relativity and information geometry.

\begin{table}[h!]
\centering
\caption{Analogy between General Relativity and Information Geometry.}
\label{tab:analogy}
\begin{tabular}{ll}
\toprule
\textbf{General Relativity} & \textbf{Information Geometry} \\
\midrule
Spacetime & Statistical Manifold \\
Event (Point in spacetime) & Probability Distribution \\
Free Particle Motion & Purely Random Process \\
& (e.g., Canonical Brownian Bridge) \\
Force & Information Constraint \\
Equivalence Principle & Equivalence Principle for Information (?) \\
\bottomrule
\end{tabular}
\end{table}

\subsection{Dual Structure in SVM}

To formulate a variational principle associated with stochastic virtual trajectories, we have to introduce dual SDEs: the forward SDE and the backward SDE. Although we do not consider the violation of metric compatibility in SVM, we introduce two mean derivatives $\uD_{\pm}$, which satisfy the following form of the stochastic integration by parts (\ref{eqn:spif}):
\begin{equation}
    \frac{\ud}{\ud t} \uE [\widetilde{X} (t) \widetilde{Y} (t)]
    = \uE [\widetilde{Y} (t ) \uD_+ \widetilde{X} (t) ]
    + \uE [\widetilde{X} (t ) \uD_- \widetilde{Y} (t) ] \, .
\end{equation}
That is, to reproduce the Leibniz rule, we have to introduce two distinct derivatives, as metric compatibility is formally violated in information geometry. 
In the stochastic calculus of variation, the two derivatives are induced by the non-differentiability of stochastic trajectories.
A deep structural analogy becomes evident when comparing this with the defining relation of dual connections in Eq.\ (\ref{eqn:non-metricity}). 
It is important to formally distinguish the two: the SVM equation is an operator relation acting on the expectation values of stochastic processes, whereas the information geometry equation is a pointwise relation defined on a statistical manifold. 
Despite this mathematical distinction, their conceptual parallelism is striking. In this analogy,
the time derivative of the expectation value $\uE[\cdot]$ in SVM plays a role analogous to the directional derivative of the metric inner product in information geometry. 
Consequently, the forward and backward derivatives, $\uD_{+}$ and $\uD_{-}$, can be viewed as the stochastic temporal counterparts to the dual affine connections, $^{(\alpha)}\nabla$ and $^{(-\alpha)}\nabla$.

In standard Riemannian geometry, metric compatibility implies a self-dual connection $(^{(\alpha)}\nabla=^{(-\alpha)}\nabla)$. 
In SVM, however, the non-differentiability of trajectories forces a splitting of the time derivative into two distinct operators, $\uD_{+}\ne \uD_{-}$. 
This explicitly breaks the self-duality in its time evolution. Viewing this through the lens of our structural analogy leads to the following interesting interpretation: the "quantum" nature of the fluctuation acts as a physical source of non-metricity in the geometry of time evolution. This further reinforces the idea that randomness is a manifestation of a richer, non-Riemannian geometric structure.

%%%%%%%%%%%%%%%%%%%%%%%%%%%%%%%%%%%%%%%%%%
\section{Summary and Concluding Remarks} \label{sec:summary}

In this review, we have established a comprehensive framework connecting Metric-Affine Gravity (MAG) and quantum mechanics through the Stochastic Variational Method (SVM). Our primary focus was to elucidate how spatial torsion, a geometric property often neglected in standard General Relativity, fundamentally alters quantum fluctuations. Contrary to the prevailing belief that torsion couples exclusively to spin degrees of freedom, we demonstrated that torsion induces measurable effects even for spinless particles.

By applying SVM in a curved geometry with totally antisymmetric torsion, we derived a non-linear Schr\"{o}dinger equation. While in Ref.\ \cite{Armin_Koide_2025}, this non-linearity arises purely from the interplay between quantum fluctuations and the torsional background, this review extends the analysis to a curved background. Here, we derived a new non-linear Schr\"{o}dinger equation containing a logarithmic non-linear term, whose coefficient is determined by the competition between the Levi-Civita part of the Ricci curvature and the torsion scalar. Crucially, this result allows us to impose stringent constraints on the magnitude of torsion based on high-precision tests of standard quantum mechanics. 
Furthermore, when applied to cosmological scales, the mathematically predicted absence of stationary states in our spatial model may theoretically harmonize with the dynamic nature of the expanding universe, which inherently prevents the formation of strictly stationary quantum states.

Moreover, this analysis points to a profound theoretical implication regarding the Geometrical Trinity of Gravity \cite{beltran2019}. While this principle establishes an equivalence within the classical framework, our results demonstrate that curvature and torsion can manifest as distinct physical entities in the quantum regime through their interaction with matter. Specifically, for spaces with negative scalar curvature, torsion cannot replicate the geometric effects of curvature induced by quantum fluctuations. 
This observation offers a concrete example of how the Geometrical Trinity breaks down beyond the classical approximation.

Furthermore, we explored the profound structural isomorphism between SVM and Information Geometry. The splitting of time derivatives ($D_+ \neq D_-$) in SVM can be interpreted as the stochastic counterpart to the dual connections ($^{(\alpha)}\nabla \neq ^{(-\alpha)}\nabla$) in information geometry, identifying quantum fluctuations as the physical origin of non-metricity.
The insights presented here open new avenues for research.

While the present analysis was restricted to a non-relativistic, trajectory-based framework where spatial torsion is treated strictly as a fixed background field, our findings lay the groundwork for several critical extensions. To achieve a more realistic application to cosmology, a necessary and intriguing next step is transitioning to the stochastic quantization of relativistic fields, such as the Klein-Gordon \cite{Koide2015} and Dirac equations. This fully covariant extension will naturally allow us to incorporate the temporal components of torsion and their full dynamics. 
In parallel, we aim to extend SVM to study torsion’s influence on the uncertainty principle \cite{G_Koide_2018,U_Gazeau_2020,Koide_2022_JSTAT} and to formulate quantum mechanics on a curved spacetime with both torsion and non-metricity, thereby exploring stochasticity from a fully geometrical perspective. Ultimately, these theoretical advancements will enable us to more rigorously investigate the implications of torsion-induced quantum fluctuations in broader cosmological contexts, including dark energy and black hole thermodynamics \cite{ONG2018217}.

%%%%%%%%%%%%%%%%%%%%%%%%%%%%%%%%%%%%%%%%%%
\vspace{6pt} 

%%%%%%%%%%%%%%%%%%%%%%%%%%%%%%%%%%%%%%%%%%
%% optional
%\supplementary{The following supporting information can be downloaded at:  \linksupplementary{s1}, Figure S1: title; Table S1: title; Video S1: title.}


% Only used for preprtints:
% \supplementary{The following supporting information can be downloaded at the website of this paper posted on \href{https://www.preprints.org/}{Preprints.org}.}

%\begin{tabularx}{\textwidth}{lll}
%\toprule
%\textbf{Name} & \textbf{Type} & \textbf{Description} \\
%\midrule
%S1 & Python script (.py) & Script of python source code used in XX \\
%S2 & Text (.txt) & Script of modelling code used to make Figure X \\
%S3 & Text (.txt) & Raw data from experiment X \\
%S4 & Video (.mp4) & Video demonstrating the hardware in use \\
%\bottomrule
%\end{tabularx}
%}

%%%%%%%%%%%%%%%%%%%%%%%%%%%%%%%%%%%%%%%%%%
\authorcontributions{Conceptualization, T.K.; methodology, T.K. and A. vdV.; validation, T.K. and A. vdV.; formal analysis, T.K. and A. vdV.; investigation, T.K. and A. vdV.; writing---original draft preparation, T.K. and A. vdV.; writing---review and editing, T.K. and A. vdV.; visualization, T.K. and A. vdV.; supervision, T.K.; project administration, T.K. All authors have read and agreed to the published version of the manuscript.}

\funding{
T.K. acknowledges the financial support by CNPq (No.\ 304504/2024-6). A.vdV. is grateful for support from the Fueck-Stiftung.
A part of this work has been done under the project INCT-Nuclear Physics and Applications (No.\ 464898/2014-5;408419/2024-5.).
}

\acknowledgments{The authors would like to thank L.\ Henrique, J.-P.\ Gazeau, G.\ Gon\c{c}alves de Matos, T.\ Kodama and K.\ Tsushima for their valuable collaboration on the research that laid the foundation for this review. During the preparation of this manuscript, the authors used Google Gemini for the purposes of grammatical review and refining the clarity of the text. The authors have reviewed and edited the output and take full responsibility for the content of this publication.}

\conflictsofinterest{The authors declare no conflicts of interest.} 

%%%%%%%%%%%%%%%%%%%%%%%%%%%%%%%%%%%%%%%%%%
%% Optional

%% Only for journal Encyclopedia
%\entrylink{The Link to this entry published on the encyclopedia platform.}

%%%%%%%%%%%%%%%%%%%%%%%%%%%%%%%%%%%%%%%%%%
%% Optional
\appendixtitles{no} % Leave argument "no" if all appendix headings stay EMPTY (then no dot is printed after "Appendix A"). If the appendix sections contain a heading then change the argument to "yes".
\appendixstart
%\appendix
%\section[\appendixname~\thesection]{}
%\subsection[\appendixname~\thesubsection]{}
%The appendix is an optional section that can contain %details and data supplemental to the main text---for %example, explanations of experimental details that would %disrupt the flow of the main text but nonetheless remain %crucial to understanding and reproducing the research %shown; figures of replicates for experiments of which %representative data are shown in the main text can be %added here if brief, or as Supplementary Data. %Mathematical proofs of results not central to the paper %can be added as an appendix.

%\begin{table}[H] 
%\caption{This is a table caption.\label{tab5}}
%\newcolumntype{C}{>{\centering\arraybackslash}X}
%\begin{tabularx}{\textwidth}{CCC}
%\toprule
%\textbf{Title 1}	& \textbf{Title 2}	& \textbf{Title 3}\\
%\midrule
%Entry 1		& Data			& Data\\
%Entry 2		& Data			& Data\\
%\bottomrule
%\end{tabularx}
%\end{table}

%\section[\appendixname~\thesection]{}
%All appendix sections must be cited in the main text. In %the appendices, Figures, Tables, etc. should be labeled, starting with ``A''---e.g., Figure A1, Figure A2, etc.

%%%%%%%%%%%%%%%%%%%%%%%%%%%%%%%%%%%%%%%%%%
\isPreprints{}{% This command is only used for ``preprints''.
\begin{adjustwidth}{-\extralength}{0cm}
} % If the paper is ``preprints'', please uncomment this parenthesis.

\reftitle{References}


% \bibliography{your_external_BibTeX_file}

%=====================================
% References, variant B: internal bibliography
%=====================================

% ACS format
\isAPAandChicago{}{%
%\begin{thebibliography}{999}
% Reference 1
%\bibitem[Author1(year)]{ref-journal}
%Author~1, T. The title of the cited article. {\em Journal Abbreviation} {\bf 2008}, {\em 10}, 142--149.
% Reference 2
%\bibitem[Author2(year)]{ref-book1}
%Author~2, L. The title of the cited contribution. In {\em The Book Title}; Editor 1, F., Editor 2, A., %Eds.; Publishing House: City, Country, 2007; pp. 32--58.
% Reference 3
%\bibitem[Author1 and Author2 (year)]{ref-book2}
%Author 1, A.; Author 2, B. \textit{Book Title}, 3rd ed.; Publisher: Publisher Location, Country, 2008; %pp. 154--196.
% Reference 4
%\bibitem[Author4(year)]{ref-unpublish}
%Author 1, A.B.; Author 2, C. Title of Unpublished Work. \textit{Abbreviated Journal Name} year, %\textit{phrase indicating stage of publication (submitted; accepted; in press)}.
% Reference 5
%\bibitem[Author8(year)]{ref-url}
%Title of Site. Available online: URL (accessed on Day Month Year).
% Reference 6
%\bibitem[Author6(year)]{ref-proceeding}
%Author 1, A.B.; Author 2, C.D.; Author 3, E.F. Title of presentation. In Proceedings of the Name of the Conference, Location of Conference, Country, Date of Conference (Day Month Year); Abstract Number %(optional), Pagination (optional).
% Reference 7
%\bibitem[Author7(year)]{ref-thesis}
%Author 1, A.B. Title of Thesis. Level of Thesis, Degree-Granting University, Location of University, %Date of Completion.
%\end{thebibliography}
%
\bibliography{sysmmetry_citedrive}
}

% If authors have biography, please use the format below
%\section*{Short Biography of Authors}
%\bio
%{\raisebox{-0.35cm}{\includegraphics[width=3.5cm,height=5.3cm,clip,keepaspectratio]{Definitions/author1.pdf}}}
%{\textbf{Firstname Lastname} Biography of first author}
%
%\bio
%{\raisebox{-0.35cm}{\includegraphics[width=3.5cm,height=5.3cm,clip,keepaspectratio]{Definitions/author2.jpg}}}
%{\textbf{Firstname Lastname} Biography of second author}

% For the MDPI journals use author-date citation, please follow the formatting guidelines on http://www.mdpi.com/authors/references
% To cite two works by the same author: \citeauthor{ref-journal-1a} (\citeyear{ref-journal-1a}, \citeyear{ref-journal-1b}). This produces: Whittaker (1967, 1975)
% To cite two works by the same author with specific pages: \citeauthor{ref-journal-3a} (\citeyear{ref-journal-3a}, p. 328; \citeyear{ref-journal-3b}, p.475). This produces: Wong (1999, p. 328; 2000, p. 475)

%%%%%%%%%%%%%%%%%%%%%%%%%%%%%%%%%%%%%%%%%%
%% for journal Sci
%\reviewreports{\\
%Reviewer 1 comments and authors’ response\\
%Reviewer 2 comments and authors’ response\\
%Reviewer 3 comments and authors’ response
%}
%%%%%%%%%%%%%%%%%%%%%%%%%%%%%%%%%%%%%%%%%%
\PublishersNote{}
\isPreprints{}{% This command is only used for ``preprints''.
\end{adjustwidth}
} % If the paper is ``preprints'', please uncomment this parenthesis.
\end{document}